\documentclass{aa} 

\usepackage{orcidlink}
\usepackage{graphicx}
\usepackage{txfonts}
\usepackage{xcolor}
\usepackage{lastpage} 
\usepackage{comment}
\usepackage{hyperref}
\hypersetup{
   colorlinks,
   filecolor=blue,
   citecolor=blue,
   linkcolor=blue,
   urlcolor=blue}

\newcommand{\lcdm}{$\Lambda$CDM}
\newcommand{\msun}{$M_\odot$}
\newcommand{\HI}{H\,{\textsc i}}
\newcommand{\HIsc}{H\,{\sc i}}
\newcommand{\ahf}{\texttt{AHF}}
\newcommand{\illustristng}{\texttt{IllustrisTNG}}

\newcommand{\tng}{\texttt{TNG50}}
\newcommand{\apostle}{\texttt{APOSTLE}}
\newcommand{\nivaria}{\texttt{NIVARIA}}
\newcommand{\nivarialg}{\texttt{NIVARIA-LG}}
\newcommand{\nihao}{\texttt{NIHAO}}
\newcommand{\nihaolg}{\texttt{NIHAO-LG}}
\newcommand{\magicc}{\texttt{MaGICC}}
\newcommand{\hestia}{\texttt{HESTIA}}
\newcommand{\auriga}{\texttt{AURIGA}}
\newcommand{\arepo}{\texttt{AREPO}}
\newcommand{\gasoline}{\texttt{GASOLINE2}}
\newcommand{\cloudy}{\texttt{CLOUDY}}
\newcommand{\pynbody}{\texttt{PYNBODY}}

\newlength{\wdth}

\begin{document} 

   \title{{\HIsc}-bearing dark galaxies predictions from constrained Local Group simulations:  how many and where to find them}

   \author{Guacimara Garc\'{i}a-Bethencourt\orcidlink{0009-0004-4798-231X}
          \inst{1}, 
          Arianna Di Cintio\orcidlink{0000-0002-9856-1943}\inst{1,2},
          Sébastien Comerón\orcidlink{0000-0002-7398-4907}\inst{1,2},     
          Elena Arjona-G\'{a}lvez\orcidlink{0000-0002-0462-7519}\inst{2,1}, \\
          Ana Contreras-Santos\orcidlink{0000-0002-3374-4626}\inst{1,2}, 
          Salvador Cardona-Barrero\orcidlink{0000-0002-9990-4055}\inst{1,2},  
           Chris B. A. Brook\orcidlink{0000-0002-0534-4115}\inst{1,2},\\
          Andrea Negri\orcidlink{0000-0003-3721-2106}\inst{1,4,5},
          Noam I. Libeskind\orcidlink{0000-0002-6406-0016}\inst{3}, and
          Alexander Knebe\orcidlink{0000-0003-4066-8307}\inst{6,7,8}
          }

   \institute{Departamento de Astrof\'{i}sica, Universidad de La Laguna, E-38200, La Laguna, Tenerife, Spain 
         \and
             Instituto de Astrof\'{i}sica de Canarias, Calle Via L\'{a}ctea s/n, E-38206 La Laguna, Tenerife, Spain 
        \and 
            Leibniz Institut für Astrophysik Potsdam (AIP), An der Sternwarte 16, D-14482, Potsdam, Germany
        \and
            Facultad de F\'{i}sica, Universidad de Sevilla, Avda. Reina Mercedes s/n, Campus de Reina Mercedes, E-41012 Sevilla, Spain
        \and
            INAF – Osservatorio di Astrofisica e Scienza dello Spazio di Bologna, Via Gobetti 93/3, I-40129 Bologna, Italy
        \and
            Departamento de F\'isica Te\'{o}rica, M\'{o}dulo 15, Facultad de Ciencias, Universidad Aut\'{o}noma de Madrid, 28049 Madrid, Spain
        \and
            Centro de Investigaci\'{o}n Avanzada en F\'isica Fundamental (CIAFF), Facultad de Ciencias, Universidad Aut\'{o}noma de Madrid, 28049 Madrid, Spain
        \and 
            International Centre for Radio Astronomy Research, University of Western Australia, 35 Stirling Highway, Crawley, Western Australia 6009, Australia
            }
   \date{Received XXX, XXXX; accepted XXX, XXXX}

 
  \abstract
  {\emph{Dark galaxies} are small, dark-matter–dominated haloes whose gas remains in hydrostatic and thermal equilibrium and has never formed stars. These systems are of particular interest because they represent a strong prediction of the {\lcdm} model. As of today, only a handful of \emph{dark galaxies} candidates have been detected so far, the most intriguing of which being Cloud-9.}
  {Using several state-of-the-art hydrodynamical simulations, we aim to predict the abundance of \emph{dark galaxies} within our Local Group, characterise their physical properties, and provide guidance for their potential observational detection.}
  {We analyse Local Group simulations with constrained initial conditions, each run with different codes, implementing different baryonic physics, feedback prescriptions, and employing two distinct values of star formation density threshold, $n_{\rm th}=0.13$ and $10\, \rm cm^{-3}$, to select samples of dark and bright galaxies harboured in haloes of similar mass.}
  {We demonstrate that \emph{dark galaxies} exist in all such simulations, though their number is larger in simulations that use a higher, more realistic $n_{\rm th}$. These galaxies, whose gas remains diffuse and never forms stars, predominantly inhabit less-concentrated, higher-spin dark matter haloes than their luminous counterparts. \emph{Dark galaxies} are typically found in low-density regions at the outskirts of the Local Group. Their formation and evolution across redshift indicate that both the dark matter and gas densities in the surroundings of \emph{dark galaxies} were consistently lower than those found around bright galaxies, making them less susceptible to interactions, mergers, or gas inflows. We estimate that up to  8 \emph{dark galaxies} should be detectable in {\HIsc} emission within $2.5\,\rm Mpc$ of the Milky Way, with the FAST radio telescope, accounting for its sky coverage and minimum {\HIsc} mass and column density.}
  {Current hydrodynamical simulations of galaxies, combined with upcoming {\HIsc} surveys, will offer a direct and powerful test of {\lcdm} through their ability to predict and measure properties of \emph{dark galaxies}  within and beyond the Local Group.}

    \titlerunning{Dark Galaxies predictions from constrained Local Group simulations}
    \authorrunning{Garc\'{i}a-Bethencourt, G., et al.}
   \maketitle
%

\section{Introduction}\label{sect: introduction}

Within the {\lcdm} model, galaxies form hierarchically through the gravitational collapse of small dark matter (DM) structures \citep{1978MNRAS.183..341White}, which then merge to form larger systems. This progressive growth of structures builds the DM haloes in which galaxies are embedded, enabling them to accrete gas that later cools and create stars. In this framework of cold DM bottom-up assembly of structures, we expect a large number of low-mass DM haloes. Some of these haloes may have a low stellar mass content or even be starless. These objects are called \emph{dark galaxies}, due to their expected DM-dominated nature (e.g. \citealt{2001MNRAS.322..658Trentham}; \citealt{2007ApJ...665L..15Kent}). 

As being such a strong prediction of the {\lcdm} model, finding \emph{dark galaxies} has been one of the main objectives of world-class radio telescopes such as the Arecibo radio telescope, through its Arecibo Legacy Fast ALFA (ALFALFA) Survey (\citealt{2005AJ....130.2598Giovanelli}; \citealt{2011AJ....142..170Haynes}), and the more recent Five-Hundred-Meter Aperture Spherical Telescope (FAST; \citealt{2024SCPMA..6719511Zhang}). 

Theoretically, it has been demonstrated that only structures with masses above a certain redshift-dependent critical mass are able to retain gas in their potential well to form stars and become luminous galaxies (e.g. \citealt{2006MNRAS.371..401Hoeft}; \citealt{2008MNRAS.390..920Okamoto}; \citealt{Benitez-Llambay2020}). This critical mass is $M_{\rm{crit}}\approx5\times 10^{9}\,M_{\odot}$  at the present day (\citealt{Benitez-Llambay2020}; \citealt{Nebrin2023}), and it depends on the time at which reionisation took place and on the mass assembly rate of each galaxy. Thus, reionisation plays an essential role in the galaxy evolution process, as it can quench and avoid star formation (SF) in small haloes, by photoheating and removing their gas (e.g. \citealt{2017MNRAS.465.3913Benitez_Llambay}). Those small haloes that are able to retain cold and dense gas to form neutral hydrogen ({\HIsc}) at their centres, but are affected by cosmic reionisation in such manner (i.e. have currently zero or few stars) have been dubbed REionisation-Limited {\HIsc} Clouds or RELHICS (\citealt{2017MNRAS.465.3913Benitez_Llambay}). These cover a halo mass range between $\sim10^{8}\,M_{\odot}$ and $\sim5\times 10^{9}\,M_{\odot}$, and are equivalent to the mini-haloes initially proposed by \cite{1986MNRAS.218P..25Rees} and \cite{1986Ap&SS.118..509Ikeuchi}. 

Due to their nature, RELHICs/\emph{dark galaxies} are challenging to observe. In fact, there are no confirmed \emph{dark galaxies}, but there are a few candidates like Cloud-9 (\citealt{Zhou2023}; \citealt{2023ApJ...956....1Benitez_Llambay}), or more recently, Candidate Dark Galaxy-2 or CDG-2 \citep{Li2025}, among others (e.g. \citealt{2005ApJ...622L..21Minchin}; \citealt{2009MNRAS.396.1096VanLoon}; \citealt{2021AJ....162..274Leisman}; \citealt{2023ApJ...944L..40Xu}; \citealt{2025SciA...11S4057Liu}). As for Cloud-9, these objects can potentially be detected as sources of {\HIsc} $21\,\rm cm$ emission without an apparent stellar counterpart, using instruments like FAST.

Cloud-9 is a potential starless \emph{dark galaxy}, detected as an {\HIsc} cloud close to M94, and appears to be devoid of a stellar component within the DESI Imaging Legacy Survey (DESI LS) surface brightness limit (\citealt{2023A&A...669A.103Martinez}). It is a gas-rich object with $M_{\rm{H\,\textsc i}}\approx10^6\,M_{\odot}$ (\citealt{2024RNAAS...8...24Karunakaran}; \citealt{Benitez-Llambay2024}) and DM halo mass of $M_{200}\sim5\times 10^9\,M_{\odot}$ (\citealt{Zhou2023}; \citealt{2023ApJ...956....1Benitez_Llambay}). Following the predictions by \cite{2017MNRAS.465.3913Benitez_Llambay}, it presents a gas temperature of $T\sim2\times 10^4\,\rm K$. Cloud-9 is estimated to be located at a distance of $d\sim4.7\,\rm Mpc$ (upper limit of $\sim10\,\rm Mpc$) and have a halo concentration of $c_{\rm{NFW}}\sim13$ (\citealt{2023ApJ...956....1Benitez_Llambay}). Moreover, a recent study by \cite{2025ApJ...993L..55Anand} confirms Cloud-9 as the first detected RELHIC. Using deep \textit{Hubble Space Telescope} (\textit{HST}) Advanced Camera for Surveys (ACS) imaging, the authors derive a new upper limit on its stellar mass of $\sim 10^{3.5}\, M_{\odot}$.

CDG-2 is also a recently discovered prospective candidate for Ultra-Diffuse/\emph{dark galaxy}, first identified by \cite{10.1214/24-AOAS1958Li2025} through the faint and diffuse emission of its population of globular clusters with \textit{HST}. It is located in the Perseus cluster, at a distance of $75\,\rm Mpc$, and has a total stellar mass of $M_{*}\approx1.2\times 10^{7}\,M_{\odot}$, and a halo mass of $M_{\rm halo}\approx2-5.7\times 10^{10}\,M_{\odot}$, which indicates that it resides in a massive DM halo (\citealt{Li2025}). 

In this context, numerical simulations developed within {\lcdm} are an important tool that can help us test results and make predictions on the nature of \emph{dark galaxies}. In particular, simulations of the Local Group are key to understand the main drivers for the formation and evolution of the galaxies residing within, as well as their environmental dependence.

For example, in the work by \cite{2017MNRAS.465.3913Benitez_Llambay}, they analysed the properties of RELHICs in a Local Group environment with the {\apostle} cosmological zoom-in simulations (\citealt{Fattahi2016}; \citealt{Sawala2016}). Their results showed that these systems harbour gas and reside beyond $500\,\rm kpc$ from the Milky Way (MW), with most of their gas in an ionised state, except for a small nearly spherical {\HIsc} core. Moreover, the gas in these systems was found to be in hydrostatic equilibrium with the underlying NFW halo (Navarro-Frenk-White; \citealt{1996ApJ...462..563NFW, 1997ApJ...490..493NFW}), and in thermal equilibrium with the cosmic UV radiation background. However, the original {\apostle} simulations of the Local Group were affected by an overproduction of stars at a fixed halo mass (\citealt{Sawala2016}), which casts doubts about the predicted number of \emph{dark galaxies} found within them.

More recently, \cite{2024ApJ...962..129Lee} used simulations from the {\illustristng} project (e.g. \citealt{2019MNRAS.490.3234Nelson}; \citealt{2019MNRAS.490.3196Pillepich}) to explore the nature of \emph{dark galaxies}. Within the {\tng} volume, they identify \emph{dark galaxies} with DM halo masses $\sim10^{9}\,h^{-1}\,M_{\odot}$ and stellar-to-total mass ratios $M_{\rm{\star}}/M_{\rm{tot}} < 10^{-4}$ at $z=0$. They find that \emph{dark galaxies} live mainly in voids and tend to be larger in size and have larger spin parameters than their luminous counterparts (see also \citealt{Jimenez2020}). However, the {\tng} simulation shares a similar issue of overcooling as the {\apostle} simulations. More specifically, below a halo mass of $10^{11}\,M_{\odot}$, most dwarf galaxies tend to overproduce stars compared to the expected abundance matching relations \citep{2013MNRAS.428.3121Moster,2014ApJ...784L..14Brook,2020A&A...634A.135Girelli}.  On top of this, \cite{2024ApJ...962..129Lee} do not require their \emph{dark galaxies} sample to explicitly contain gas, making thus difficult to use their predictions in light of upcoming {\HIsc} surveys.
The tendency of \emph{dark galaxies} to live in low density regions has also been recently explored observationally in \cite{2025ApJS..279...38Kwon}, who study the properties of a selection of prospective \emph{dark galaxy} candidates from the ALFALFA Survey (\citealt{2018ApJ...861...49Haynes}).

In this work, we move forward by studying the number, distribution and characteristics of \emph{dark galaxies} arising in different sets of zoom-in Local Group simulations, and offer concrete predictions on their detectability with radio telescopes. For this analysis, we identify \emph{dark galaxies} in four simulated Local Group runs employing constrained initial conditions: three from the {\hestia} simulations  \citep{2020MNRAS.498.2968Libeskind}, and one from the new {\nivarialg} simulation \citep{ContrerasSantos_etal}. In particular, the three high-resolution {\hestia} runs reach down to $1.5\times10^{5}\,M_{\odot}$ in DM particle mass and $220\,\rm pc$ in spatial resolution. In contrast, the {\tng} simulations have a DM particle mass resolution of $4.5\times10^{5}\,M_{\odot}$ and a softening length of $330\,\rm pc$. The {\nivarialg} simulations, although lower in resolution than {\hestia}, closely follow the expected abundance matching relation at all masses, lending confidence to the reliability of their \emph{dark galaxy} predictions.

We explore the properties of \emph{dark galaxies} within these simulations which are run with different codes, galaxy formation models, baryonic physics, feedback implementations, resolutions, and SF density thresholds, with the aim of validating whether \emph{dark galaxies} can be found across different simulation setups. 
Our goal is to characterise the main properties of {\HI}-bearing \emph{dark galaxies}  in Local Group-like environments and to study their evolution over time, shedding light on their nature and uncovering details of their origin and formation pathways. To better understand these objects, we compare our \emph{dark galaxy} sample with a sample of bright galaxies, i.e. luminous galaxies within the same halo mass range. Additionally, we treat Cloud-9 as a reference for our simulated \emph{dark galaxies} and test our results against the {\HIsc} detection limits of the FAST radio telescope. 

We describe the simulation sets in detail in Section~\ref{sect: simulations}. In Section~\ref{sect: results}, we present the selection criteria and resulting samples of dark and bright galaxies from each simulation, including properties of the underlying DM haloes, gas content, environmental effects, and {\HIsc} detectability with FAST. Section~\ref{sect: discussion} provides a discussion of the results, and Section~\ref{sect: conclusions} summarises the main conclusions.

\section{Simulations}\label{sect: simulations}
We employ two different sets of cosmological hydrodynamical simulations that reproduce the properties of the Local Group, including two massive central haloes that are analogues of the MW and M31. These simulation suites consist of the three high-resolution simulations from the {\hestia} project (\citealt{2020MNRAS.498.2968Libeskind}) and a lower-resolution simulation named {\nivarialg} \citep{ContrerasSantos_etal}. Both sets are zoom-in simulations run from an original cosmological box of size $100\,h^{-1}\,\rm Mpc$. The simulations adopt the parameters of a {\lcdm} Planck cosmology (\citealt{2014A&A...571A..16Planck}), i.e. $\Omega_{\rm m} = 0.307$, $\Omega_{\rm b} = 0.048$, $\Omega_{\Lambda} = 0.693$, $\sigma_{8} = 0.8288$, and $H_{0} = 100\,h\,\rm km\,s^{-1}\,\rm Mpc^{-1}$, where $h = 0.6777$. They also employ constrained initial conditions derived from observations of the peculiar velocity field in the CosmicFlows-2 catalogue (\citealt{2013AJ....146...86Tully}), which are constructed to replicate the main large-scale structures of the Local Volume at $z = 0$ (e.g. a Local Group-like pair, a Virgo Cluster-like halo; \citealt{1991ApJ...380L...5Hoffman}; \citealt{1999ApJ...520..413Zaroubi}). For both suites, the properties of the haloes are derived using the Amiga Halo Finder\footnote{\urlstyle{rm}\url{http://popia.ft.uam.es/AHF/}}({\ahf}; \citealt{gill2004, knollmann2009}), which defines haloes as overdensities of DM with $\Delta \sim 200$ times the critical density of the Universe at $z = 0$, $\rho_{\rm c} = 3H_{0}^{2}/(8\pi G)$. We refer hereafter to $M_{200}$ and $R_{200}$, given by the {\ahf} code, as $M_{\rm{halo}}$ and $R_{\rm{vir}}$, respectively. Likewise, we refer to these haloes as ‘galaxies’ in this paper. {\ahf} is also used to trace haloes across snapshots, i.e. to construct a merger tree that connects haloes between consecutive time steps in the simulations. For the analysis and post-processing of the simulations, we make use of the {\pynbody}\footnote{\urlstyle{rm}\url{https://pynbody.readthedocs.io/latest/}} package \citep{pynbody}. 

Despite the parallels between the two sets of simulations, they were run with different codes, subgrid physics, feedback prescriptions, and SF models. These differences make the {\hestia} and {\nivarialg} simulations ideal for comparison and for assessing the robustness of our results. A more detailed description of the simulations is provided below.

\subsection{HESTIA simulations}
We use the three highest-resolution simulations from the `High-resolutions Environmental Simulations of The Immediate Area' {\hestia} project, whose initial conditions seeds are named 09\_18, 17\_11, and 37\_11 \citep{2020MNRAS.498.2968Libeskind}.
{\hestia} is the successor of the original `Constrained Local UniversE Simulations'
{\texttt{CLUES} project \footnote{\urlstyle{rm}\url{www.clues-project.org}} (\citealt{2010arXiv1005.2687Gottloeber}; \citealt{2010MNRAS.401.1889Libeskind}; \citealt{2016MNRAS.458..900Carlesi}; \citealt{2016MNRAS.455.2078Sorce}). The high-resolution region consists in two spheres of $3.7\,\rm Mpc$ ($2.5\,h^{-1}\, \rm{Mpc}$) of radius centred on the two main haloes of each run at $z = 0$. The particle masses resolution for DM, gas, and stars are $m_{\rm{DM}}= 1.5\times 10^{5}\,M_{\odot}$, $m_{\rm{gas}}= 2.2\times 10^{4}\,M_{\odot}$, and $m_{\rm{star}}= 2.0\times 10^{4}\,M_{\odot}$, respectively. The softening length is $\epsilon= 220\,\rm pc$. 
These simulations were run with the {\auriga} galaxy formation model (\citealt{2017MNRAS.467..179Grand}) employing the {\arepo} \textit{N}-body moving-mesh code (\citealt{2010MNRAS.401..791Springel}; \citealt{2016MNRAS.455.1134Pakmor}; \citealt{2020ApJS..248...32Weinberger}). The formation model incorporates magnetic fields and super-massive black holes (SMBHs) physics. A spatially uniform UV background is set as an ionising source of radiation, which completes the reionisation epoch at $z \sim 6$ (\citealt{2013MNRAS.436.3031Vogelsberger}). Gas cooling is implemented for primordial gas and metals (\citealt{2013MNRAS.436.3031Vogelsberger}). 
Gas is converted into stars stochastically following a Kennicutt-Schmidt relation (\citealt{1959ApJ...129..243Smidth}; \citealt{1998ApJ...498..541Kennicutt}) when its density reaches values larger than $n_{\rm{th}}=0.13\,\rm{cm}^{-3}$. Gas that satisfies this density criterion follows a two-phase subgrid model that distinguishes between cold and hot components of the interstellar medium (ISM; \citealt{2003MNRAS.339..289Springel}). An effective equation of state is applied to maintain pressure equilibrium in the star-forming gas (e.g. \citealt{2017MNRAS.467..179Grand}).

    \begin{figure}
    \centering
    \includegraphics[width=9cm]{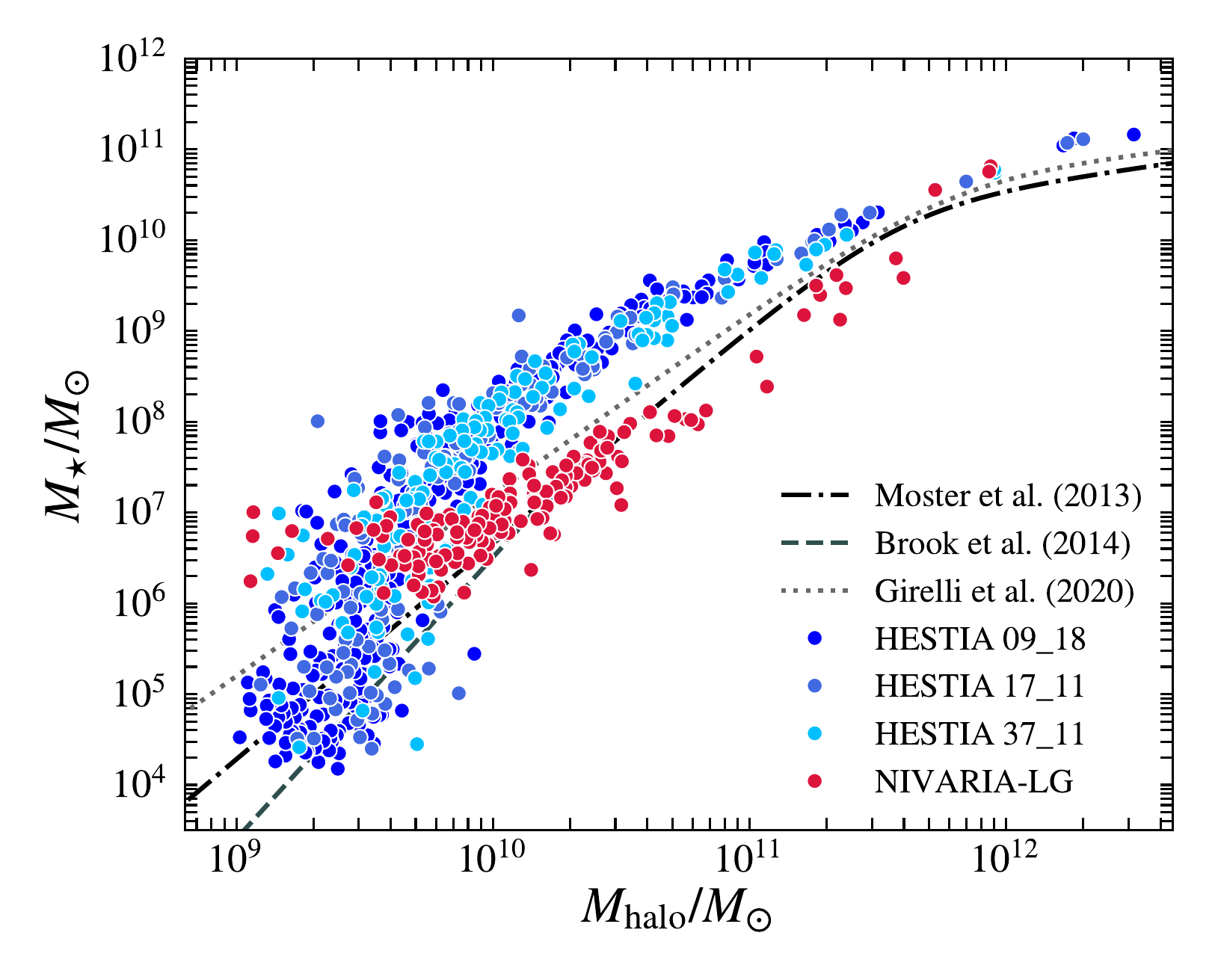}
    \caption{$M_{\star}$–$M_{\rm halo}$ relation for isolated galaxies within our simulated Local Group analogues, containing gas and stars and with DM halo masses between $10^{9}\,M_{\odot}$ and $10^{12.5}\,M_{\odot}$. Galaxies from the {\hestia} simulations are coloured in shades of blue, and galaxies from {\nivarialg}  in red. The $M_{\star}-M_{\rm halo}$ relations from \cite{2013MNRAS.428.3121Moster}, \cite{2014ApJ...784L..14Brook}, and \cite{2020A&A...634A.135Girelli} are shown in dotted-dashed, dashed, and dotted lines, respectively.} 
    \label{fig: allmhalo_mstar}
    \end{figure}

\subsection{NIVARIA-LG simulations}
The `Numerical InVestigation of dwARf galaxies Including AGN feedback in the Local Group', {\nivarialg} project (\citealt{ContrerasSantos_etal}), is a new set of simulations whose initial conditions are drawn from one of the {\hestia} intermediate-resolution runs, and are based on the galaxy formation model from the {\nihao} (\citealt{2015MNRAS.454...83Wang}) and {\magicc} (\citealt{stinson2012}; \citealt{2012MNRAS.424.1275Brook}) projects, which have been shown to follow well observed galaxy scaling relations (e.g. \citealt{2012MNRAS.424.1275Brook}; \citealt{2015MNRAS.454...83Wang}; \citealt{2016MNRAS.463L..69Maccio};  \citealt{2017MNRAS.467.4937Dutton}; \citealt{2018MNRAS.473.4392Santos-Santos}). 

The simulations comprise two analogue runs: one in which a SMBH is seeded at the centre of haloes exceeding $M_{\rm halo}\sim10^{10}M_{\odot}$, and another in which SMBH seeding is disabled. In this work, we use the run without the SMBH implementation.
The  cosmological initial conditions used to generate the Local Group environment in {\nivarialg} are drawn from one of the {\hestia} intermediate-resolution runs, following an approach similar to that adopted in \cite{arora22}. The zoom-in region of {\nivarialg} consists of a sphere of radius $5\, h^{-1}\rm Mpc$ centred on the Local Group. The particle masses for DM, gas, and stars are $m_{\rm DM} = 1.6\times 10^{6}\, M_{\odot}$, $m_{\rm gas} = 3\times 10^{5}\, M_{\odot}$, $m_{\rm star} = 6\times 10^{4}\, M_{\odot}$, respectively. The minimum gravitational softening lengths are $\epsilon_{\rm DM} = 860\rm\, pc$ for DM and $\epsilon_{\rm gas} = 488\rm\, pc$ for gas.
This cosmological hydrodynamical simulation is run with a modified version of the Smoothed Particle Hydrodynamics (SPH) code {\gasoline} \citep{2017MNRAS.471.2357Wadsley}. The BH implementation is described in \cite{2019MNRAS.487.5476Blank}, which is also the one used in \cite{2024MNRAS.533.1463Waterval}, while the chemistry evolution is the same as {\nihaolg} (\citealt{arora22}). Gas and metal-line cooling are implemented for H, He, and several metal species (\citealt{shen2010}), using {\cloudy} \citep{Ferland_1998}. A uniform UV background \citep{haardt_madau1996} provides ionising radiation and heating of the medium, modelling reionisation, which is completed at $z \sim 6$. The density threshold for SF is set to $n_{\rm th} = 10\rm\, cm^{-3}$, consistent with the densities of giant molecular clouds. Gas is also required to have a temperature below $T_{\rm th} = 1.5\times10^{4}\rm\, K$. SF follows a \cite{2003PASP..115..763Chabrier} initial mass function and the Kennicutt–Schmidt relation (\citealt{1959ApJ...129..243Smidth}; \citealt{1998ApJ...498..541Kennicutt}). Supernovae types Ia and II, as well as early stellar feedback \citep{stinson2012}, are included to inject energy and metals into the surrounding medium in both runs.

An important difference arising from the distinct subgrid models and feedback implementations in the two simulation suites is that {\hestia} tends to form more stars than expected from observations at fixed halo mass. As a result, {\hestia} galaxies lie systematically above the empirical stellar-to-halo mass ($M_{\star}$–$M_{\rm halo}$) relation, as seen in Fig.~\ref{fig: allmhalo_mstar}. This overproduction of stars, associated with the \citet{2003MNRAS.339..289Springel} model, has been noted previously (e.g. \citealt{2013ApJ...763L..41Benitez}; \citealt{2025A&A...699A.301Arjona}), and is particularly pronounced for haloes with masses between $10^{10}\, M_{\odot}$ and $10^{11}\, M_{\odot}$. Conversely, owing to their different SF criteria and feedback prescriptions, {\nivarialg} galaxies follow the empirical relations more closely. In particular, when compared with observational estimates, the {\nivarialg} simulations reproduce both the stellar-to-halo mass relation (e.g. the observational points in \citealt{2017MNRAS.467.2019Read} and the empirically derived abundance-matching relations shown in Fig.~\ref{fig: allmhalo_mstar}) and the cold gas–to–stellar mass relation (see Fig.~3 of \citealt{arora22}, based on a companion {\nivaria}-like simulation, with  similar resolution and feedback scheme).
This contrast allows us to bracket the expected number of \emph{dark galaxies} providing upper and lower limits from the two simulation suites as a reference for observational expectations.

\section{Results}\label{sect: results}

\subsection{Dark galaxy selection sample}\label{subsect: selection}
We select samples of \emph{dark galaxies} and bright galaxies from each simulation at $z = 0$. Both samples are defined within a sphere of $2.5\,\rm Mpc$ in radius centred on the MW analogues in order to ensure the search within the highest-resolution regions of the Local Group. We select only those galaxies with a DM halo mass in the [$10^{9}-10^{10}$] {\msun} range and a fraction of high-resolution DM particles larger than 0.98. The upper limit of $10^{10}\, M_{\odot}$ is chosen because we do not expect any galaxy above this halo mass to remain dark, while the lower limit of $10^{9}\, M_{\odot}$ is dictated by the resolution and the necessity of having both dark and bright galaxies within the same halo mass range.

   \begin{table} 
      \caption[]{Sample of galaxies resulting from the selection described in Section~\ref{subsect: selection} for each simulation. The first column lists the label assigned to each simulation. The second column provides the total mass contained within $2.5\,\rm Mpc$ of the corresponding MW analogue. The third column reports the total number of galaxies identified within $2.5\,\rm Mpc$ of the MW and with halo masses between $10^9$ and $10^{10}\,M_{\odot}$ using the selection criteria. The fourth column shows the number of bright galaxies, while the fifth column lists the number of \emph{dark galaxies} in each simulation. The final column displays the number of \emph{dark galaxies} from the previous column that are completely starless. Values in brackets indicate galaxies that are satellites of others (i.e. systems that are not isolated).}
         \label{tab: sample}
         \centering
         \tabcolsep=0.14cm
          \begin{tabular}{l c c c c c}
            \hline\hline
            Sim.      &  $M_{\rm{tot}} 
             $ & $n_{\rm{total}}$ & $n_{\rm{bright}}$ & $n_{\rm{dark}}$ & $n_{\rm{starless}}$ \\
            &$(10^{12} \, M_{\odot})$\\
            \hline
            \footnotesize{{\hestia} 09\_18} & 13.03 & 187 (27) & 84 (18) & 32 (0) & 17 \\
            \footnotesize{{\hestia} 17\_11} & 10.77 & 139 (14) & 56 (11) & 22 (0) & 11 \\
            \footnotesize{{\hestia} 37\_11} & 7.67  & 133 (25) & 53 (19) & 1  (1) & 0 \\
            \footnotesize{{\nivarialg}}     & 5.51  & 81  (9)  & 15 (4) & 59 (5)   & 59\\
            \hline
         \end{tabular}   
   \end{table}

Here, we define \emph{dark galaxies} as those haloes that, following the conditions described above, have $\leq 10$ stellar particles, and $\geq 10$ gas particles. This is $\lesssim 10^{5}\,M_{\odot}$ in stellar mass, and $\gtrsim 10^5-10^{6}\,M_{\odot}$ in gas mass, depending on the resolution of the simulations. The number of gas particles was chosen to guarantee that the predicted \emph{dark galaxies} preserve sufficient gas to remain detectable with radio telescopes. On the other hand, we select a sample of bright galaxies as those haloes containing  more than 10 stellar particles. In this case, we do not restrict to a minimum  number of gas particles. Hence, some of these galaxies could be gas-free at $z = 0$ (and hence non-star forming),  due to gas exhaustion after SF, ram pressure stripping (\citealt{1972ApJ...176....1GunnGott}), or other processes such as interaction with the cosmic web (e.g. \citealt{2013ApJ...763L..41Benitez}).

    \begin{figure}
    \centering
    \includegraphics[width=9cm]{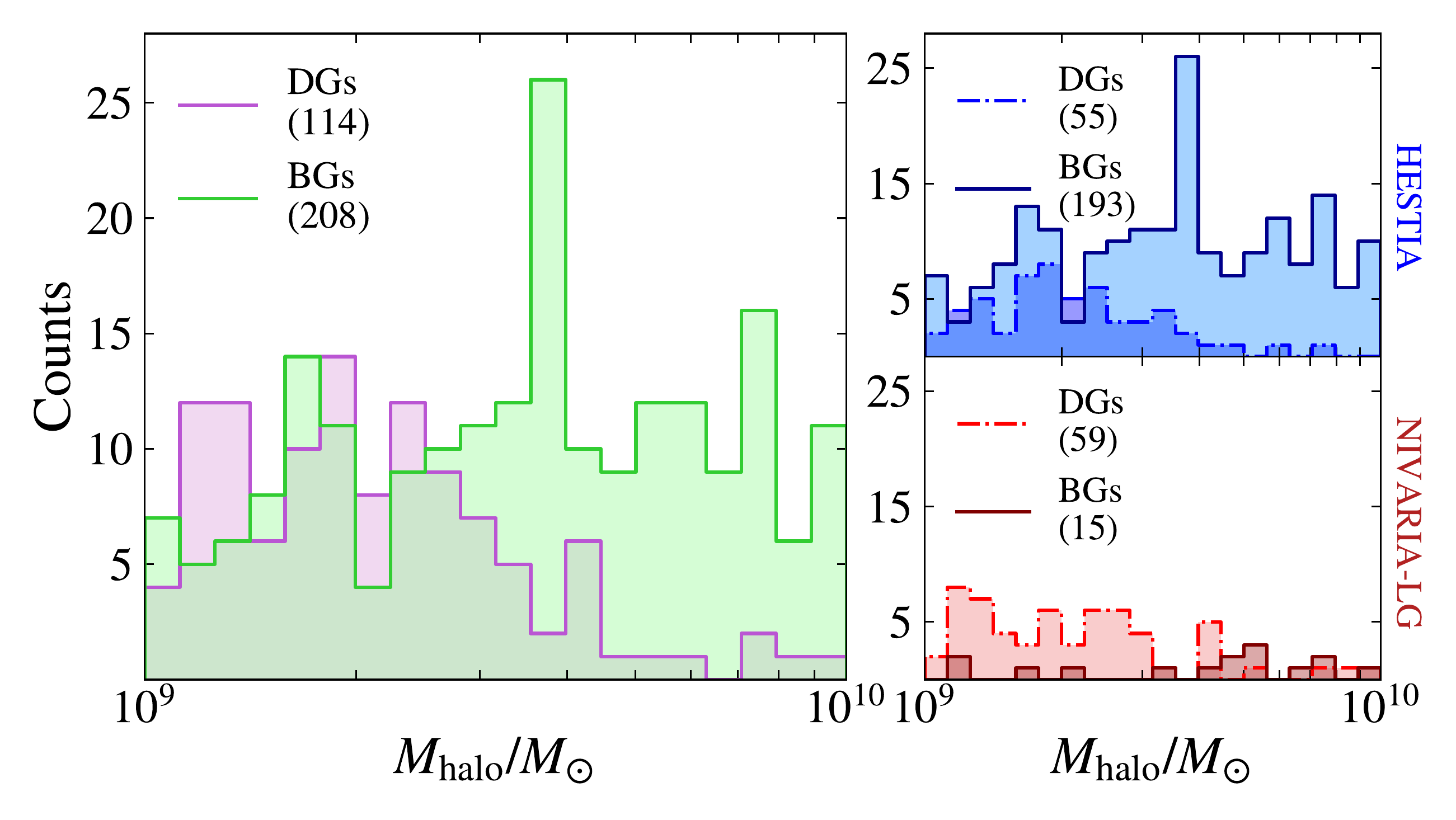}
    \caption{Number of galaxies from the four simulations as a function of the DM halo mass (\textit{left panel}). The total sample of \emph{dark galaxies} is shown in magenta and the total sample of bright galaxies is shown in green. The \textit{right} panels show the same distributions separately for {\hestia} simulations in blue (\textit{top}) and for {\nivarialg} simulations in red (\textit{bottom}), representing dark and bright galaxies with dotted-dashed and solid lines, respectively.} 
          \label{fig: mhalo_hist}
    \end{figure}

Table~\ref{tab: sample} shows the number of galaxies found after applying the selection cuts described above. From the second and third columns, we note that, as expected, the total number of galaxies depends on the total mass of the simulation within our designated region, with the most massive simulation containing the largest number of galaxies. However, the number of bright galaxies in the {\hestia} simulations is much larger than the number of \emph{dark galaxies}, whereas the opposite is true for {\nivarialg}. This difference is likely a consequence of the different SF density thresholds employed in the two simulation sets. Therefore, we do not apply any mass rescaling to the predicted numbers of dark and bright galaxies based on the total Local Group mass, as these numbers depend not only on mass but also on the underlying subgrid physics models of each simulation. Finally, the number of satellites in each sample is indicated in brackets, i.e., haloes belonging to a central or host galaxy. In both {\hestia} and {\nivarialg}, a much higher fraction of bright galaxies are satellites compared to \emph{dark galaxies}, suggesting that bright galaxies tend to reside closer to other systems and may experience more interactions.

    \begin{figure}
    \centering 
    \includegraphics[scale=0.32, trim={0cm 3.3cm 0cm 0.7cm}, clip] {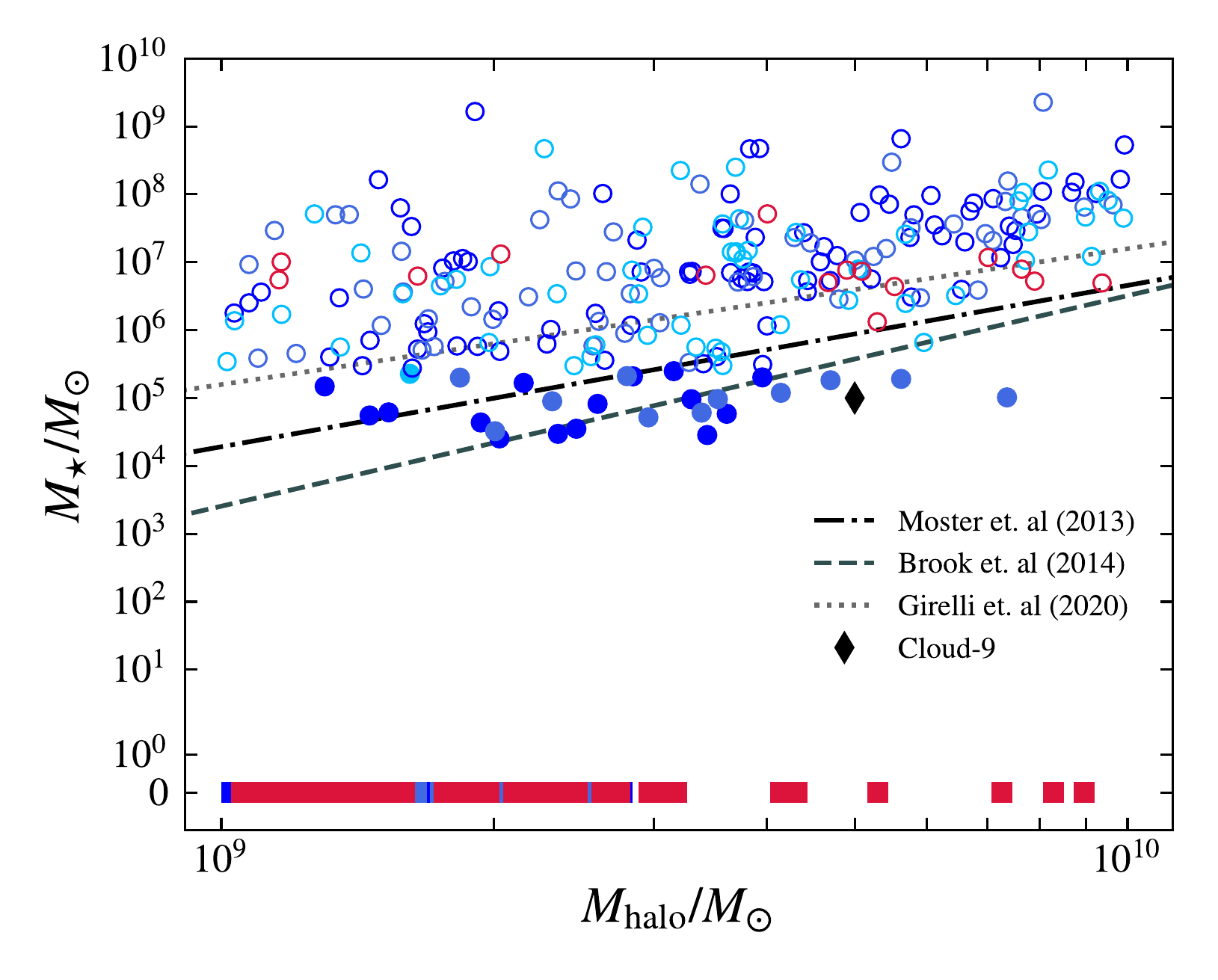}
    \includegraphics[scale=0.32, trim={0cm 3.3cm 0cm 0.7cm}, clip]{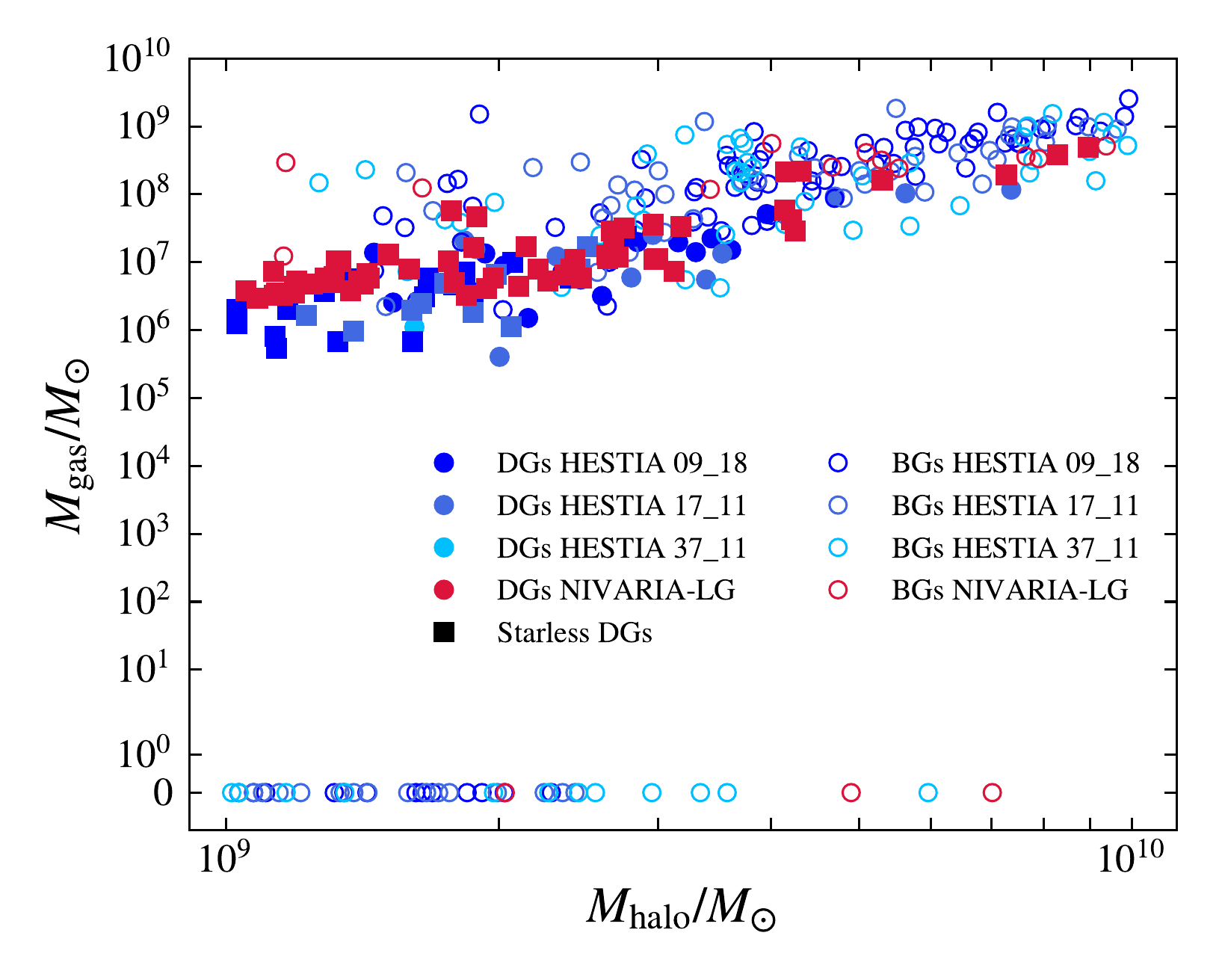}
    \includegraphics[scale=0.32, trim={0cm 0cm 0cm 0.7cm}, clip]{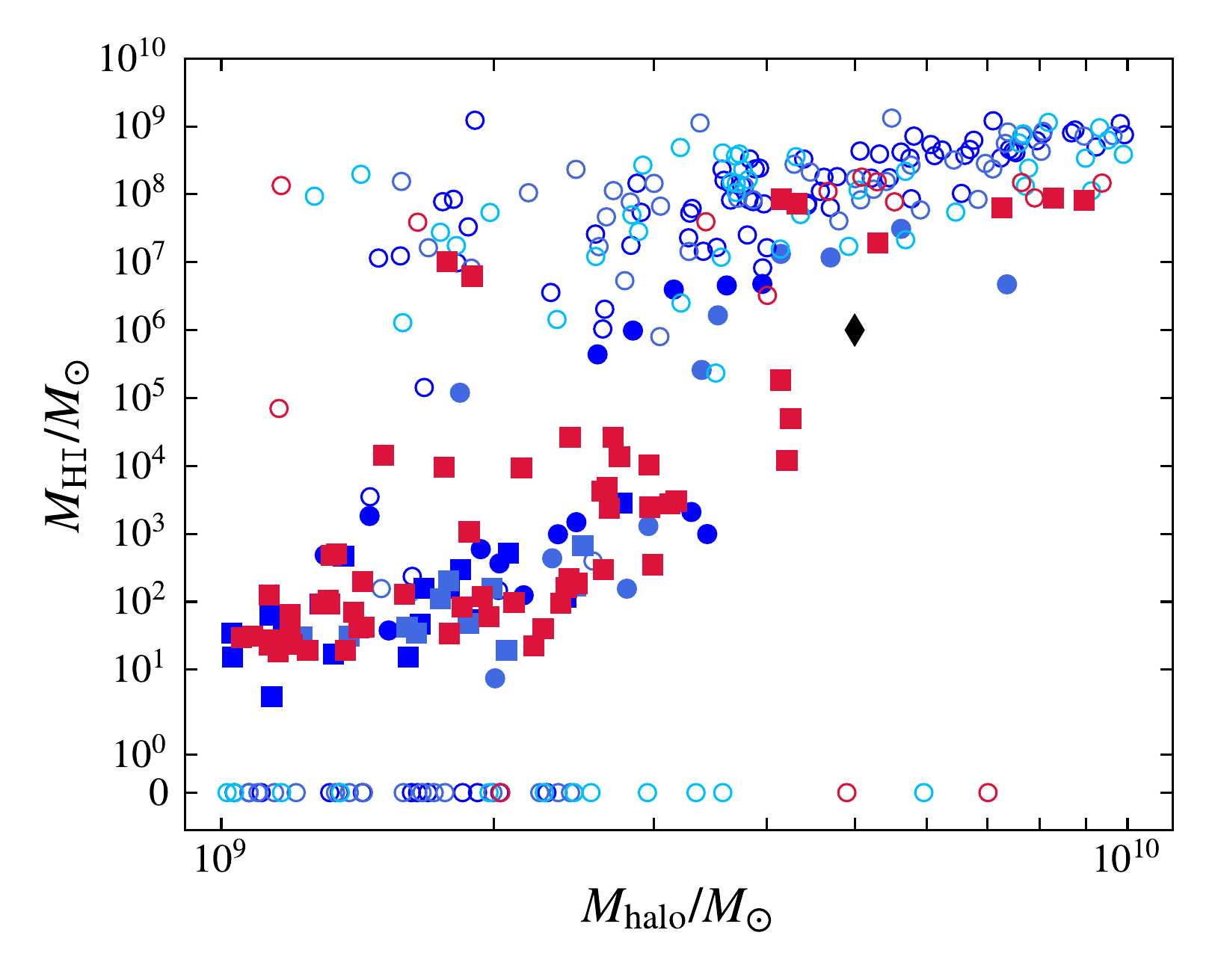}
    \caption{$M_{\star}$–$M_{\rm halo}$ relation (\textit{top}), gas-to-halo mass relation (\textit{centre}), and H\,{\sc i}-to-halo mass relation (\textit{bottom}), for dark (filled markers) and bright (empty markers) galaxies at $z = 0$. Galaxies from the {\hestia} simulations are coloured in shades of blue and galaxies from {\nivarialg} are in red. Circles indicate \emph{dark galaxies} with stars, whereas squares indicate  starless \emph{dark galaxies}. In the \textit{top} panel, the $M_{\star}$–$M_{\rm halo}$ relations from \cite{2013MNRAS.428.3121Moster}, \cite{2014ApJ...784L..14Brook}, and \cite{2020A&A...634A.135Girelli} are shown in dotted-dashed, dashed, and dotted lines, respectively. We include Cloud-9 as a black diamond in the \textit{top} and \textit{bottom} panels.}
              \label{fig: mhalo_mstar_mgas}%
    \end{figure}

It is important to note that the selection cuts for \emph{dark galaxies} could be affected by the different particle resolutions of {\hestia} and {\nivarialg}. In Fig.~\ref{fig: allmhalo_mstar}, for halo masses below $\sim 5 \times 10^{9}\,M_{\odot}$, {\hestia} simulations can form less massive isolated galaxies than {\nivarialg}. This difference can be attributed to the combination of lower mass resolution and a more restrictive SF criterion adopted in {\nivarialg}. If {\nivarialg} had the same particle resolution as {\hestia} while preserving its SF density threshold, we would expect a slightly smaller number of \emph{dark galaxies}. A higher resolution would lower the effective halo mass threshold required for SF, increasing the number of haloes that host stellar particles at a fixed halo mass. In this sense, our predictions should be regarded as an upper limit to the number of \emph{dark galaxies} expected in the Local Group.

To visualise the distribution of the sample, Fig.~\ref{fig: mhalo_hist} shows the number of dark and bright galaxies as a function of DM halo mass, combining all the simulation sets in the left panel, from {\hestia} only in the top right panel, and from {\nivarialg} only in the bottom right panel. The values in brackets in the legends correspond to the total number of galaxies for each case. Bright galaxies dominate the more massive halo region of the plot, whereas \emph{dark galaxies} are more abundant in the lower-mass end of the distribution. This tendency is consistent with the galaxy sample selection from \cite{2024ApJ...962..129Lee}.

In the top panel of Fig.~\ref{fig: mhalo_mstar_mgas}, we show the $M_{\star}$–$M_{\rm halo}$ relation for dark and bright galaxies at $z = 0$ (referred as DGs and BGs in legends hereafter, respectively), in tandem with the semi-empirical relations from \cite{2013MNRAS.428.3121Moster}, \cite{2014ApJ...784L..14Brook}, and \cite{2020A&A...634A.135Girelli}. Similarly, we show the relation between the gas and the halo masses in the central panel, and the relation between {\HIsc} gas  and halo masses at the bottom (see Sec.~\ref{subsect: HIFAST} for  details on how {\HIsc} is computed in each simulation). Cloud-9 is included in the top and bottom panels for reference, assuming $M_{200} \sim 5\times 10^9 M_{\odot}$ (\citealt{2023ApJ...956....1Benitez_Llambay}), an upper stellar mass limit of $10^5 M_{\odot}$ (\citealt{Zhou2023}; \citealt{Benitez-Llambay2024}), and $M_{\rm \HI} \approx 10^6 M_{\odot}$ (\citealt{Benitez-Llambay2024}). It falls within the massive end of our \emph{dark galaxies} region. 

In the top panel of the figure, we see a clear separation between dark and bright galaxies in the \textit{y}-axis, around $10^{5}\,M_{\odot}$ in stellar mass and a large number of starless galaxies, especially for {\nivarialg}, which we show as squares. More specifically, the number of starless \emph{dark galaxies} in our sample is 17 for the {\hestia} 09\_18 run, 11 for {\hestia} 17\_11, 0 for {\hestia} 37\_11, and 59 for the {\nivarialg} simulation (see Table~\ref{tab: sample}). That is, the whole sample of {\nivarialg} \emph{dark galaxies} is composed of completely starless objects. 

On the other hand, in the central panel, the total gas mass exhibits a clear correlation with DM halo mass, and the dark and bright galaxy samples appear broadly similar in this respect. Some bright galaxies, however, have no gas content and are therefore non-star-forming at the present time. In contrast, in the bottom panel of Fig.~\ref{fig: mhalo_mstar_mgas}, the {\HIsc} gas content differs between the two samples: only the most massive \emph{dark galaxies} have {\HIsc} masses comparable to their bright counterparts, i.e., $M_{\rm \HI} > 10^5\, M_{\odot}$ for $M_{\rm halo} > 10^{9.5}\, M_{\odot}$.

\subsection{Dark matter haloes properties}

In this section, we tackle the evolution of DM halo mass and its relation to reionisation, as well as intrinsic properties of the haloes, like the concentration and the spin parameter. 

In order to analyse the impact of reionisation, we need to estimate the evolution with time of the minimum critical halo mass needed for gas to collapse and form stars. Following the scheme given in \cite{Benitez-Llambay2020}, before reionisation the critical DM halo mass corresponds to the one above which atomic hydrogen can cool efficiently to form stars:
   \begin{equation}
      M_{\rm{H}}^{z} \sim (4 \times 10^7 \; M_{\odot}) \left(\dfrac{1 + \it z}{11}\right)^{-3/2},  
   \end{equation}

\noindent On the other hand, after reionisation sets off, the critical DM halo mass for which gas can collapse within haloes is:
   \begin{equation}
      M_{\rm{crit}}^{z} \sim (10^{10} \; M_{\odot}) \left(\dfrac{\it T_{\rm{b}}}{3.2 \times 10^4 \; \rm{K}}\right)^{3/2} (1 + \it z)^{-3/2},  
   \end{equation}
where $T_{\rm{b}}$ refers to the virial temperature of the halo.

In Fig.~\ref{fig: masses_z}, we show such critical mass as a function of $z$. We compare  the median halo mass evolution of our dark (dotted-dashed lines) and bright (solid lines) galaxies for {\hestia} simulations in blue and {\nivarialg} in red, together with theoretical predictions. Here, the cyan dashed line represents $M_{\rm{crit}}^{z}$ for a fixed halo with $T_{\rm{b}} = 2 \times 10^4\,\rm K$, and the orange line represents $M_{\rm{H}}^{z}$. By combining the two relations prior and after reionisation, \cite{Benitez-Llambay2020} show the global critical mass for which gas can collapse across redshift, represented in Fig.~\ref{fig: masses_z} by the black dotted-dashed line, leading to a minimum halo of $M_{200} \sim 5 \times 10^9\,M_{\odot}$ at $z=0$. Thus, a halo that lies below the black dotted-dashed line should remain completely dark (see also \citealt{2023MNRAS.519.1425Pereira-Wilson}). This plot displays a substantial difference between dark and bright galaxies, where the former tend to lie close or below the critical mass, whilst the latter are well above the black line at all $z$, indicating their luminous nature. 

We can observe that \emph{dark galaxies} generally form later-on than bright galaxies, being their half-mass accretion time $z_{\rm dark}=2.0$ and  $z_{\rm dark}=1.7$ for \texttt{NIVARIA-LG} and  \texttt{HESTIA}, respectively, while the formation time of  the corresponding bright galaxies is between $z_{\rm bright}=3.2$ and $z_{\rm bright}=2.6$.
This result is in agreement with findings presented in \cite{2024ApJ...962..129Lee} (their Fig.~5 and discussion therein). In Fig.~\ref{fig: masses_z} it is also clear that bright galaxies, which are overall more massive than \emph{dark galaxies} (see Fig.~\ref{fig: mhalo_hist}), were already more massive prior to reionisation, making them  less vulnerable to the effects of a ionising background. 
The later formation times of \emph{dark galaxies} should be reflected in the concentration parameters of their DM haloes, as these systems formed in a less dense Universe. 

    \begin{figure}
    \centering
    \includegraphics[width=9cm]{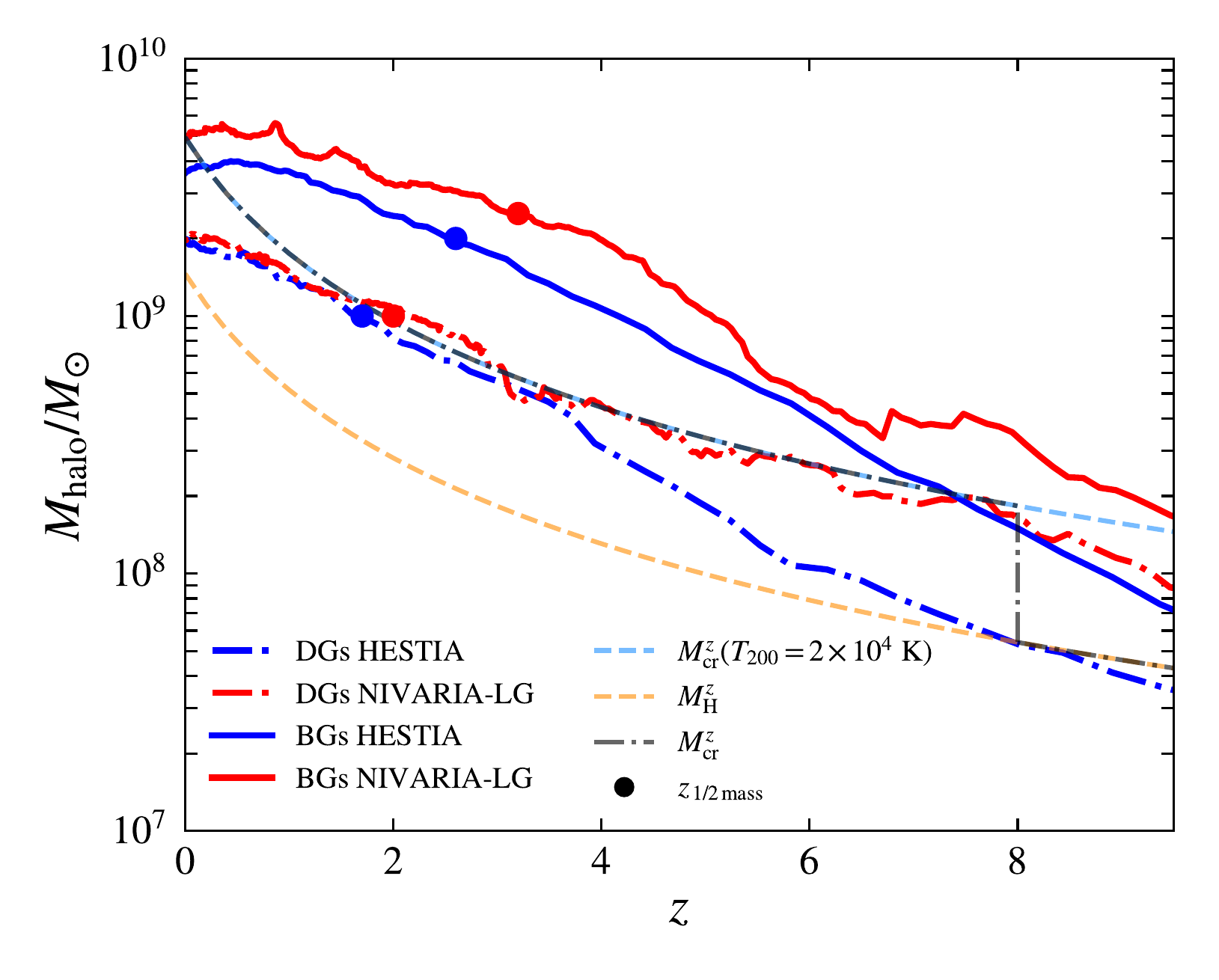}
    \caption{Evolution of the median DM halo masses of \emph{dark galaxies} (dotted-dashed lines) and bright galaxies (solid lines) as a function of redshift. Red lines represent the median of all galaxies of the {\nivarialg} simulation, whilst blue lines correspond to the median of all galaxies for the {\hestia} runs. The red and blue dots indicate the redshifts at which the galaxies reach half of their final halo mass. Following the scheme from \cite{Benitez-Llambay2020}, we show with the grey dotted-dashed line the critical mass needed for gas to collapse. In addition, the cyan dashed line shows the critical halo mass for a fixed virial temperature of $T_{\rm{b}} = 2 \times 10^4\,\rm K$. The orange dashed line represents the halo mass above which atomic hydrogen cooling is efficient.} 
          \label{fig: masses_z}%
    \end{figure}   

    \begin{figure}
    \centering
    \includegraphics[width=9cm]{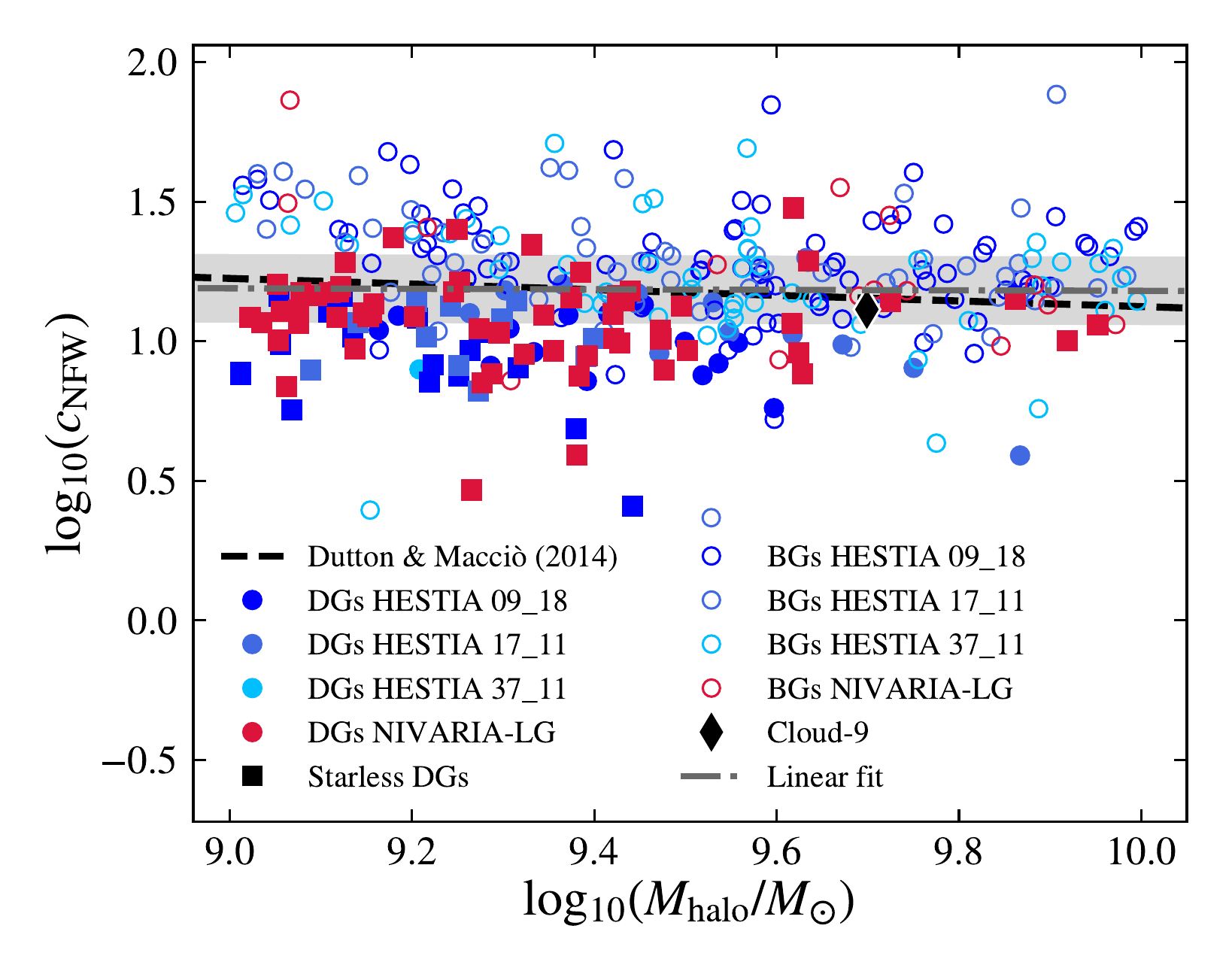}
    \caption{Concentration-halo mass relation for dark (filled markers) and bright (empty markers) galaxies at $z = 0$, colour-coded as in Fig.~\ref{fig: mhalo_mstar_mgas}. The black dashed line is the $c_{200}$ relation drawn from \cite{2014MNRAS.441.3359DuttonMaccio}, and the grey dotted-dashed line represents a lineal fit of both the dark and bright sample together, i.e. $\rm{log}_{10}(\textit{c}_{\rm{NFW}}) = -0.01\, \rm{log}_{10}(\textit{M}_{\rm{halo}}/\textit{M}_{\odot})+1.27$. The shaded region shows the $3\sigma$ scatter of both relations.}
          \label{fig: mhalo_cNFW}%
    \end{figure}

    \begin{figure}
    \centering
    \includegraphics[width=8.5cm]{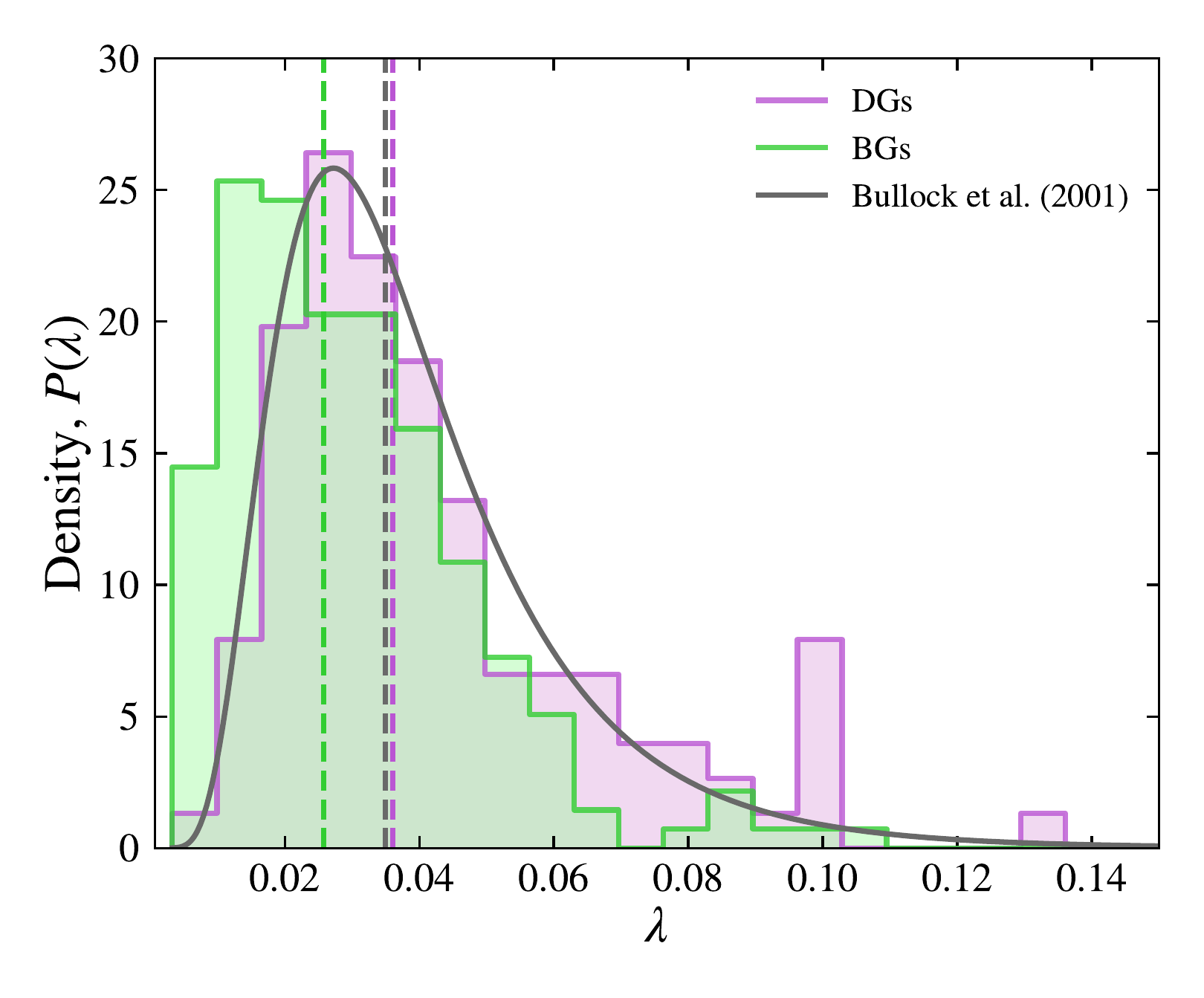}
    \caption{Probability density distribution of the spin parameter of the DM halo, $P(\lambda)$, for \emph{dark galaxies} in magenta and bright galaxies in green in all four simulations. The vertical dashed lines indicate the median values for each distribution, being $\lambda_{\rm{dark}} = 0.036 \pm 0.003$ for \emph{dark galaxies}, and $\lambda_{\rm{bright}} = 0.026 \pm 0.002$ for bright galaxies. The solid grey line represents the log-normal fiducial distribution from \cite{2001ApJ...555..240Bullock}, and the dashed grey line indicates its median, i.e. $\lambda_{0} = 0.035 \pm 0.005$.} 
          \label{fig: spinparameter}%
    \end{figure}

    \begin{figure*}
   \centering
    \includegraphics[scale=0.188, trim={0cm 4.3cm 0.9cm 0cm}, clip]{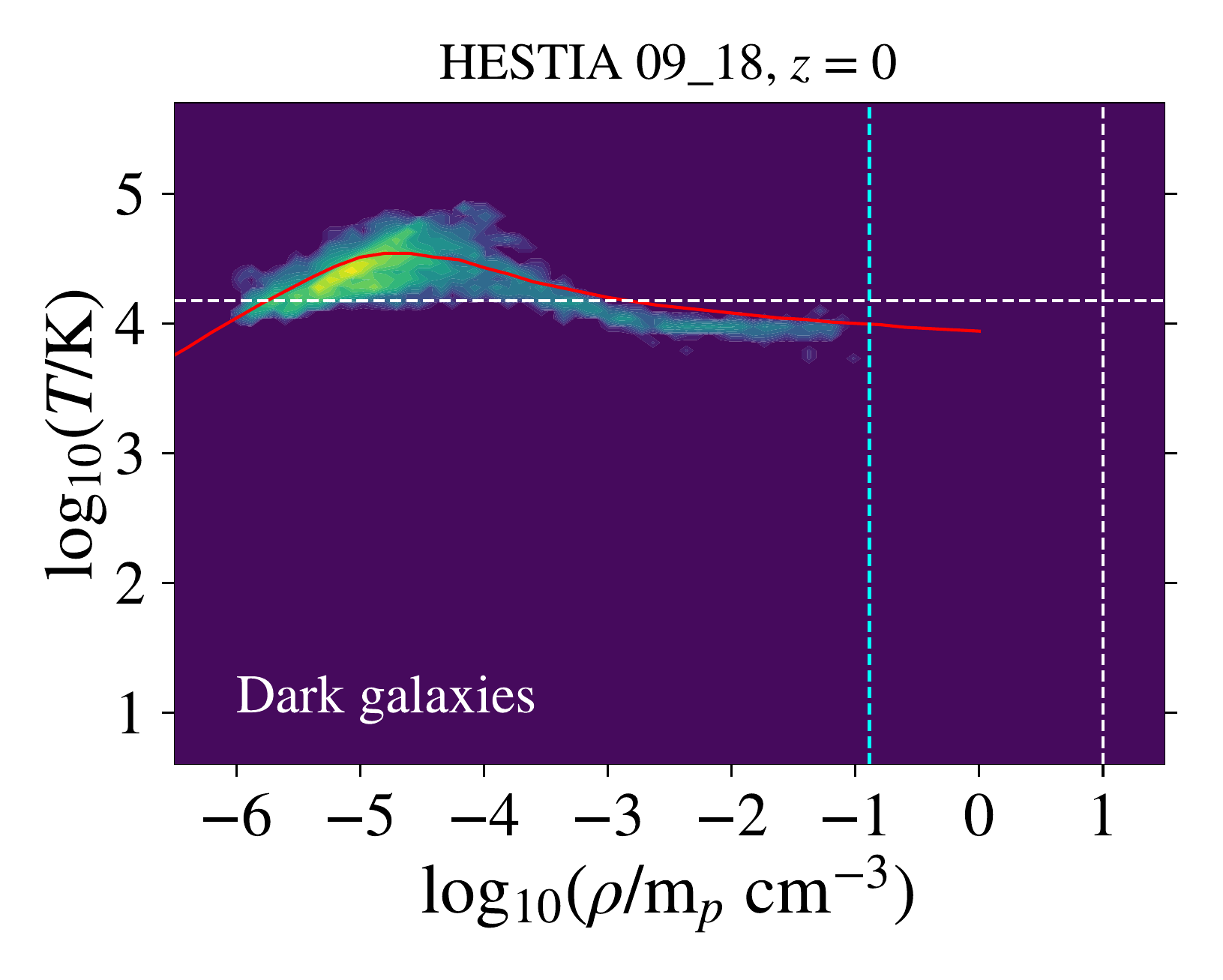} 
    \includegraphics[scale=0.188, trim={3.6cm 4.3cm 0.9cm 0cm}, clip]{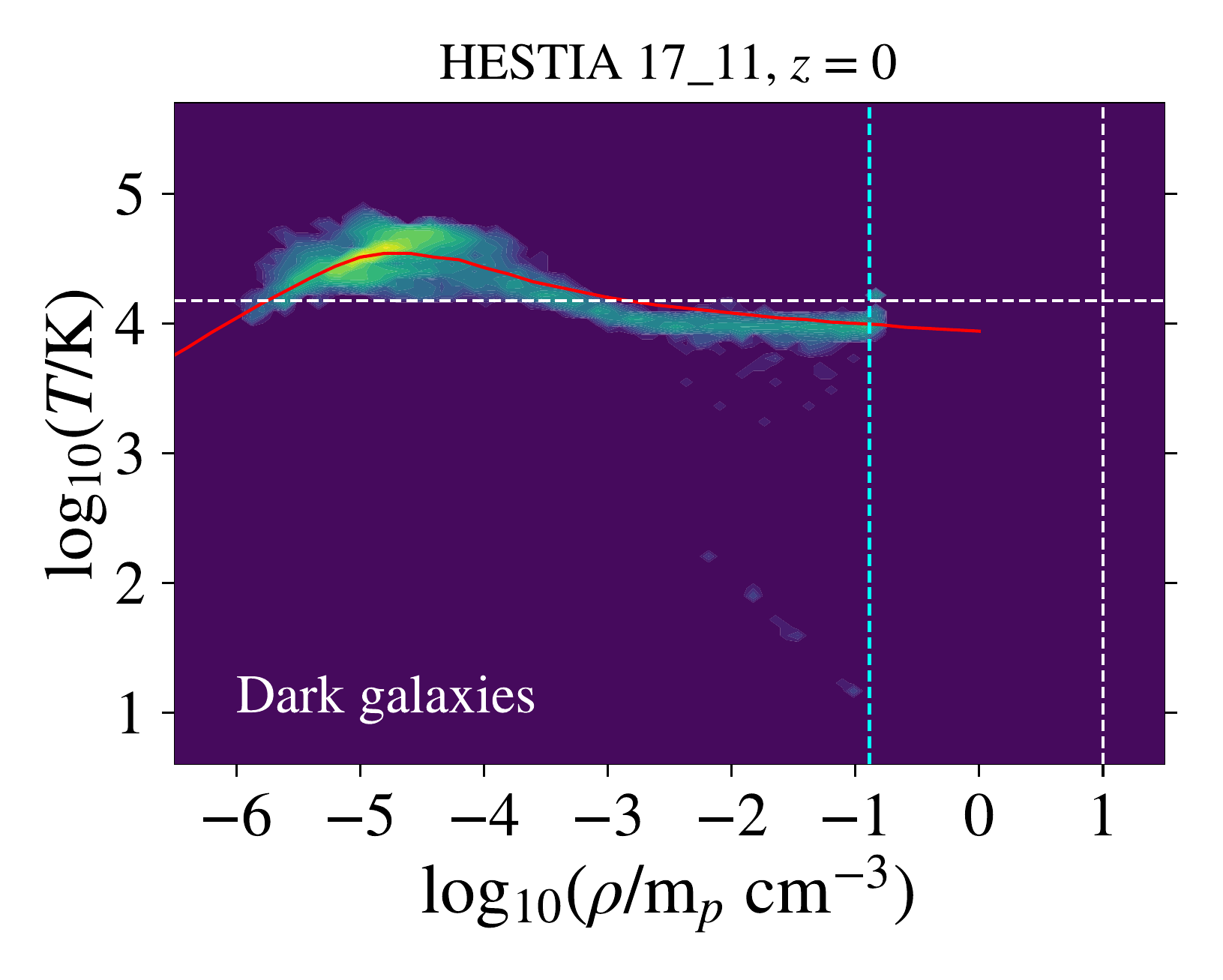}
    \includegraphics[scale=0.188, trim={3.6cm 4.3cm 0.9cm 0cm}, clip]{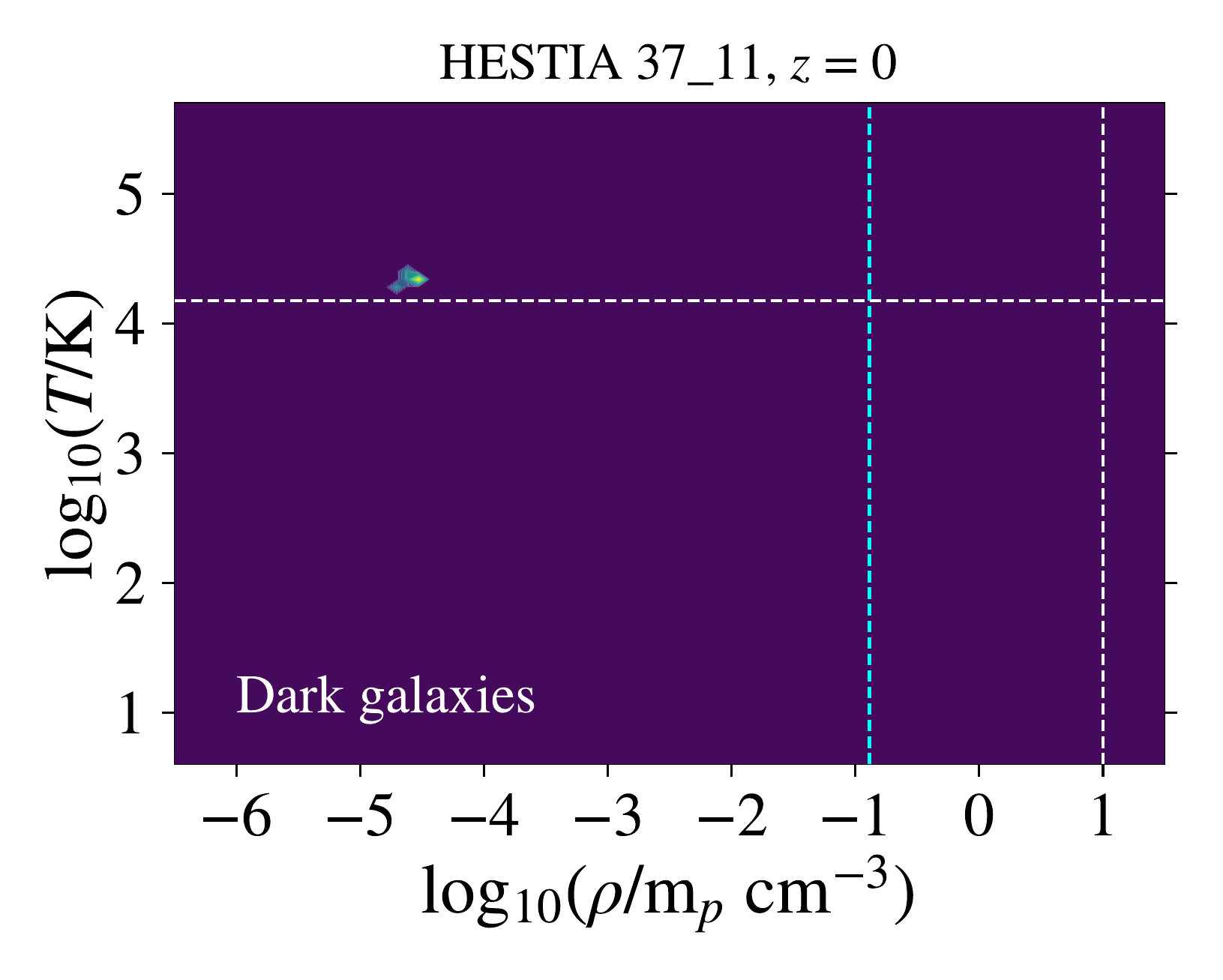} 
    \includegraphics[scale=0.188, trim={3.6cm 4.3cm 0.9cm 0cm}, clip]{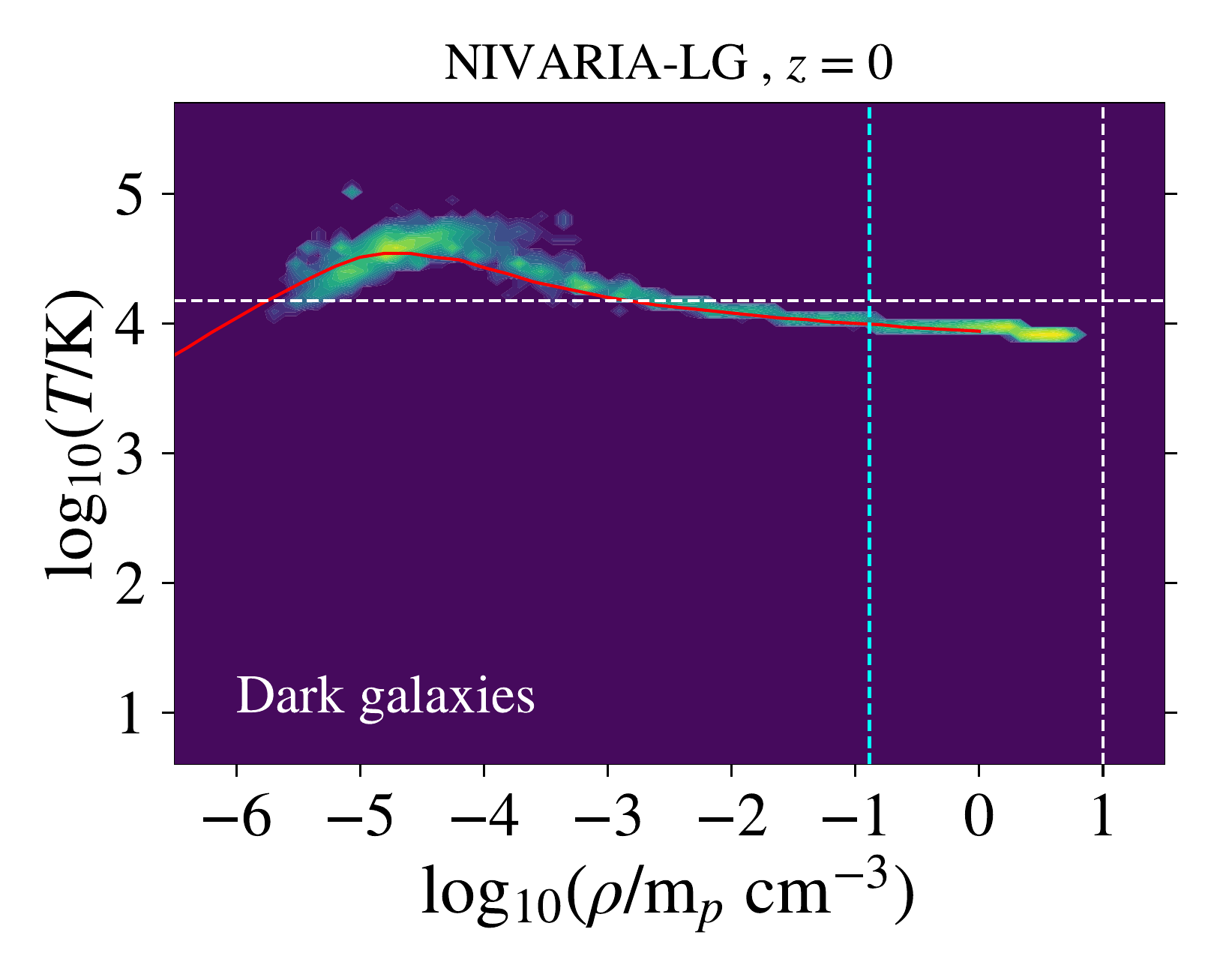}\\
    \includegraphics[scale=0.187, trim={0cm 1cm 0.9cm 2.0cm}, clip]{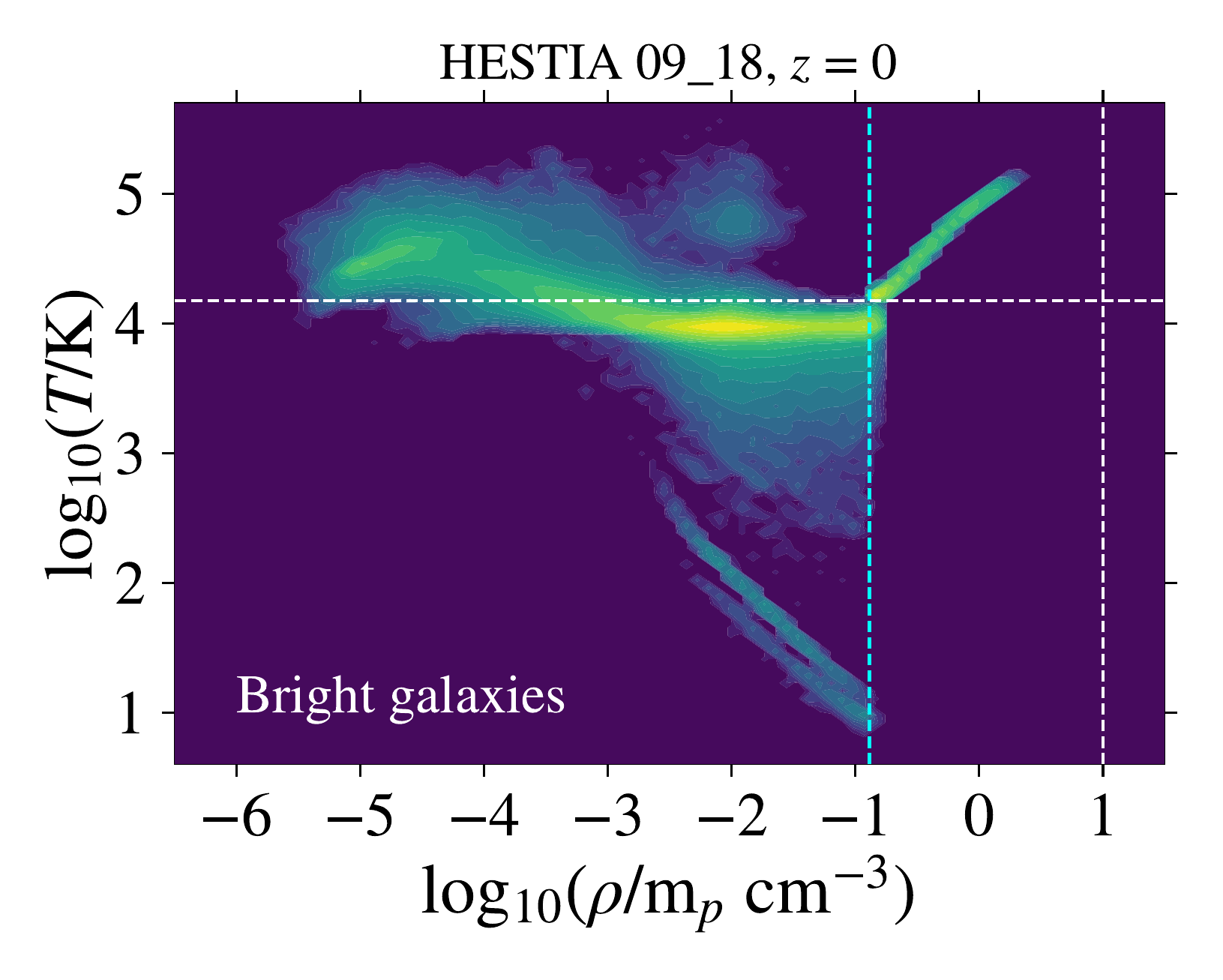} 
    \includegraphics[scale=0.1878, trim={3.6cm 1cm 0.9cm 2.0cm}, clip]{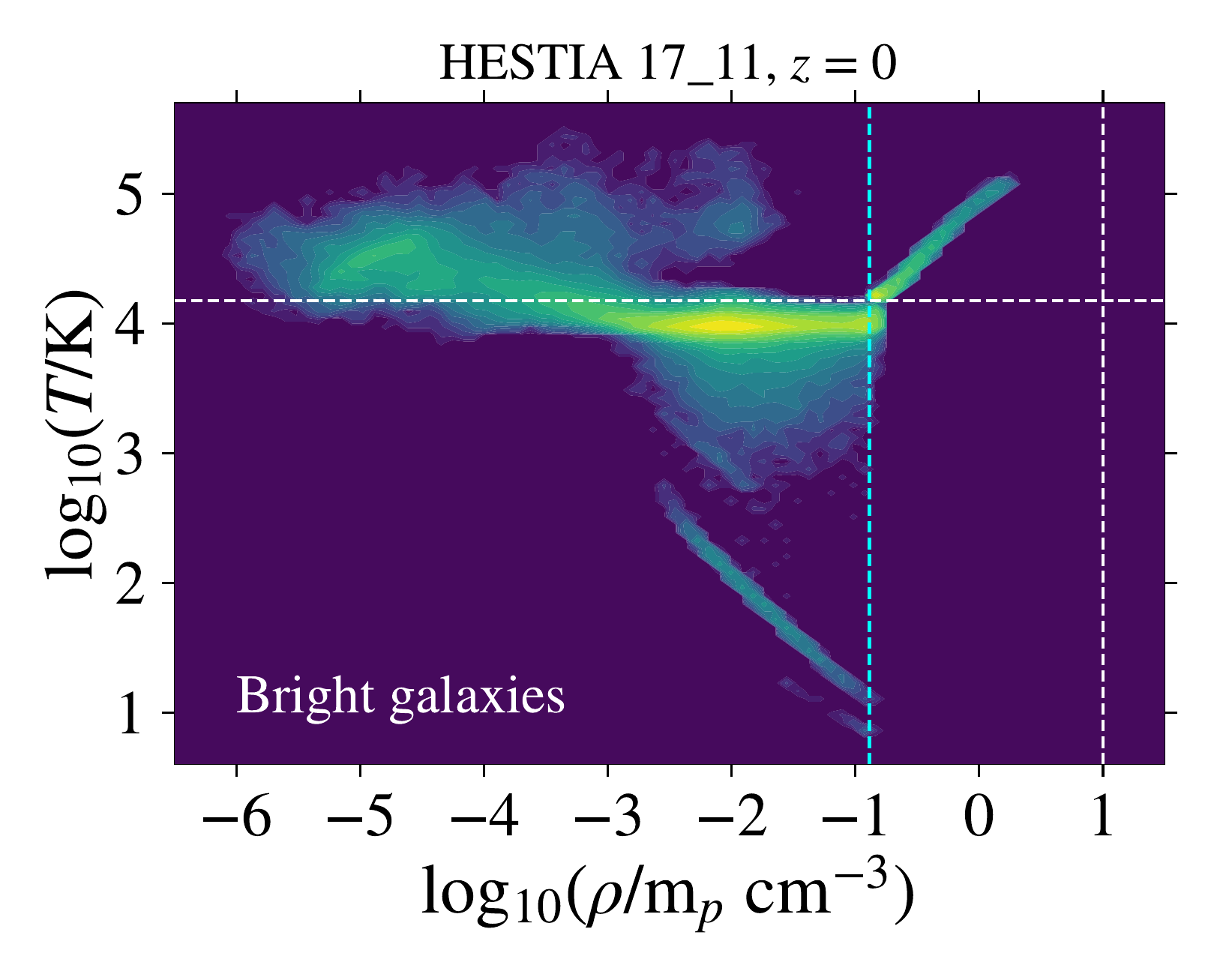}
    \includegraphics[scale=0.1878, trim={3.6cm 1cm 0.9cm 2.0cm}, clip]{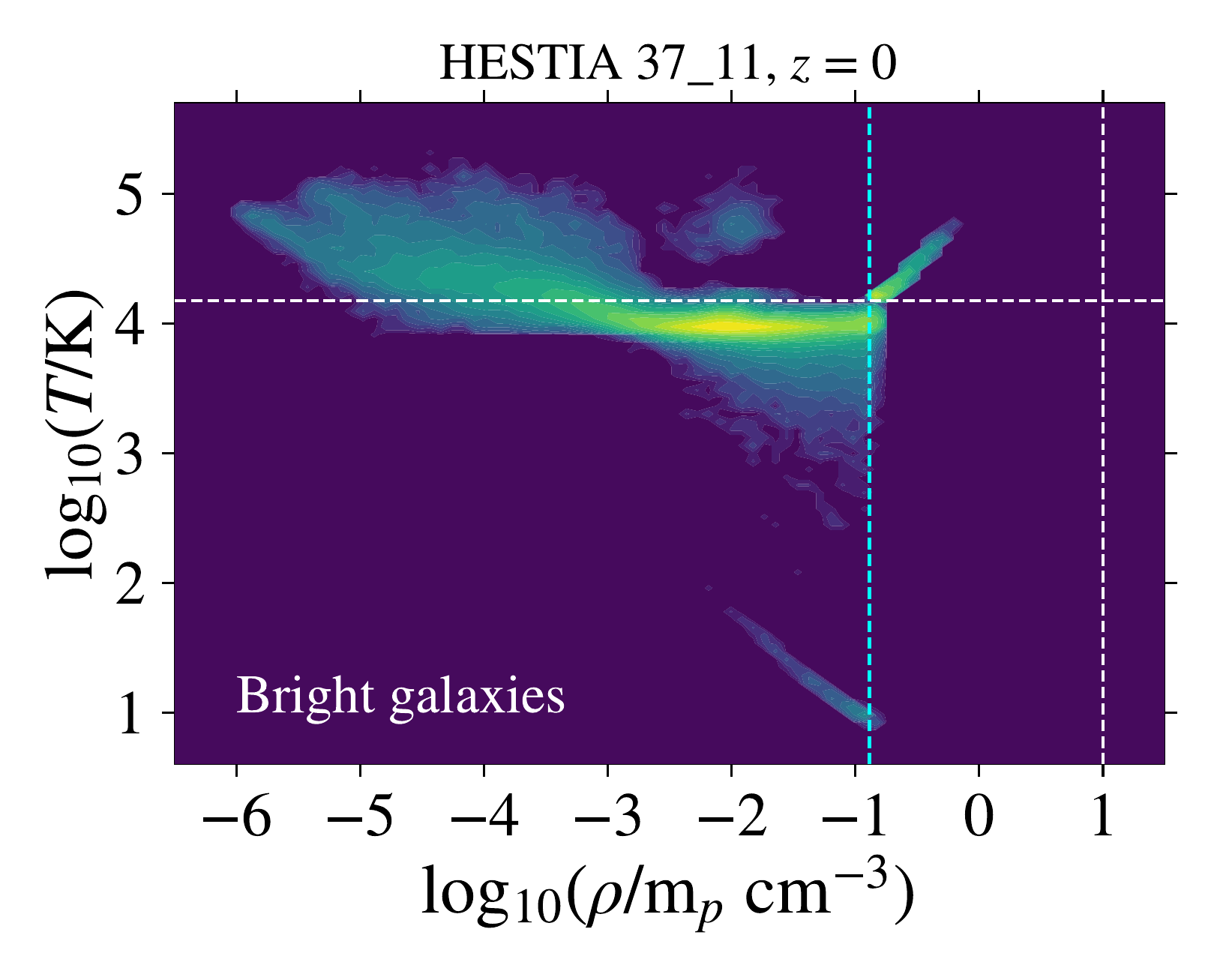} 
    \includegraphics[scale=0.187, trim={3.6cm 1cm 0.9cm 2.0cm}, clip]{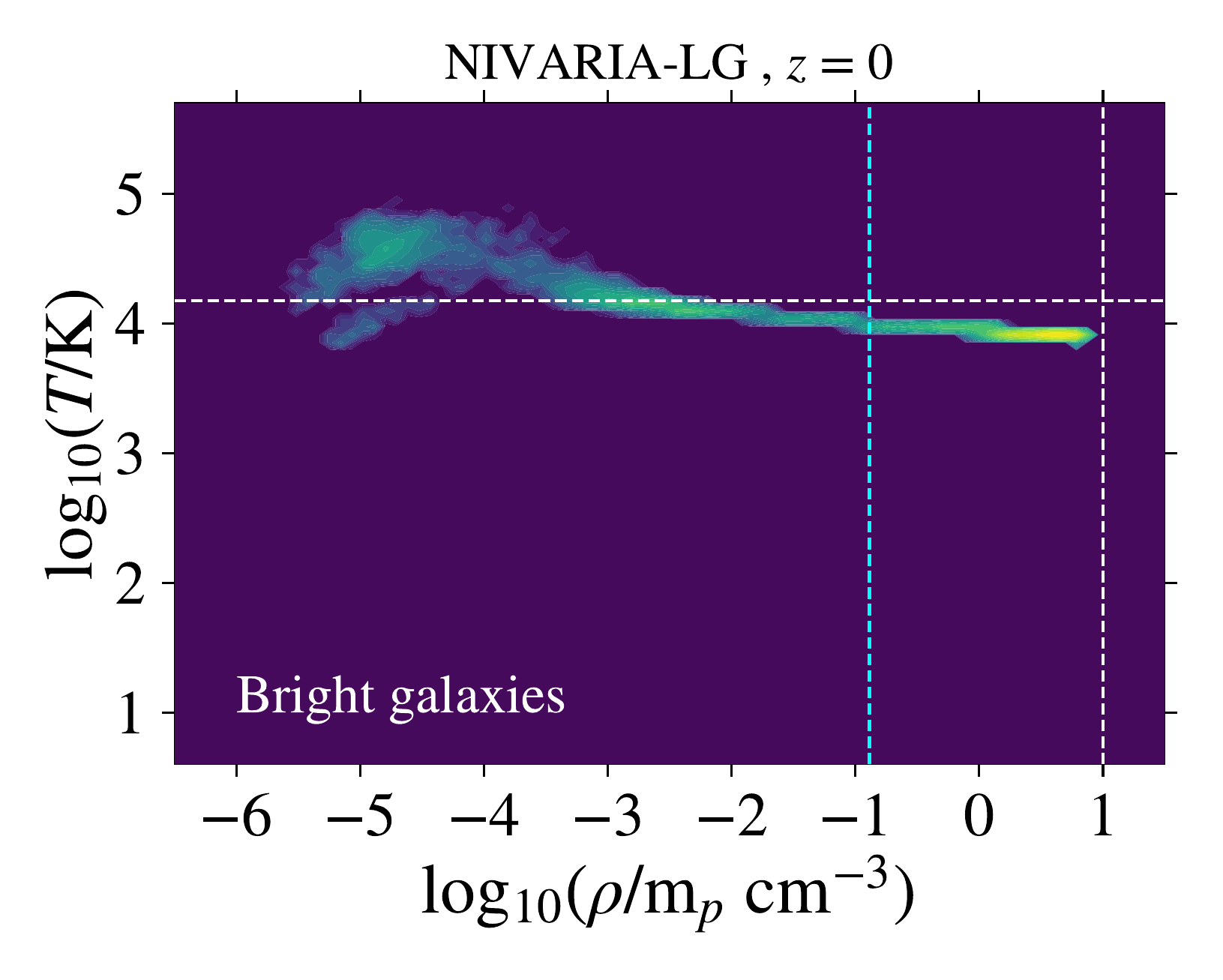}
   \caption{Temperature-density diagram for gas particles in all dark (\textit{top row}) and all bright (\textit{bottom row}) galaxies. Each column corresponds to a different simulation, from \textit{left} to \textit{right} {\hestia} 09\_18, 17\_11, 37\_11, and {\nivarialg}. The horizontal dashed line indicates the threshold temperature below which gas is  able to form stars in {\nivarialg}, whilst the vertical cyan and white lines show the corresponding density thresholds to set off SF in {\hestia} ($n_{\rm th}=0.13\,\rm cm^{-3}$) and {\nivarialg} ($n_{\rm th}=10\, \rm cm^{-3}$), respectively. The red continuous line in the $T-\rho$ diagrams of \emph{dark galaxies} correspond to the predicted temperature-density relation expected in RELHICs, as derived in \cite{2017MNRAS.465.3913Benitez_Llambay}. Note that the \emph{dark galaxy} histogram corresponding to {\hestia} 37\_11 is rather sparse, as only one object of this type is found in that simulation. } 
              \label{fig: gas_tempdens}%
    \end{figure*}
    
Figure~\ref{fig: mhalo_cNFW} shows the relation between the halo mass and the concentration corresponding to DM haloes with a NFW profile at $z = 0$, where the black dashed line represents the $c_{200}-M_{\rm 200}$ relation from \citealt{2014MNRAS.441.3359DuttonMaccio}, following a Planck cosmology. In this figure, the grey dotted-dashed line corresponds to the lineal fit of the data (both dark and bright galaxies), and the shaded region is the $3\sigma$ scatter between this fit and the \cite{2014MNRAS.441.3359DuttonMaccio} relation. As expected, we see that the \emph{dark galaxy} sample has less concentrated DM haloes than bright galaxies at a fixed halo mass. Quantitatively, \emph{dark galaxies} have a median value of $c_{\rm{NFWdark}} = 11.1 \pm 0.5$, and bright galaxies a median value of $c_{\rm{NFWbright}} = 18.6 \pm 0.5$. Cloud-9, also added in the plot, lies close to the \cite{2014MNRAS.441.3359DuttonMaccio} relation.
    
We then study the spin parameter of the DM haloes in both samples, shown in Fig.~\ref{fig: spinparameter}. The spin parameter, $\lambda$, is defined by the following expression (\citealt{2001ApJ...555..240Bullock}): 

   \begin{equation}
      \lambda = \dfrac{|\boldsymbol{J}|}{\sqrt{2}MRV} \,,  
   \end{equation}

\noindent where $\boldsymbol{J}$ is the angular momentum of a sphere with mass $M$ and radius $R$, and $V$ is the circular velocity at $R$.

We  compare the resulting distributions of the spin parameters with the log-normal distribution from \cite{2001ApJ...555..240Bullock} as a reference.
\emph{Dark galaxies} (coloured in magenta) show larger spin parameters than bright galaxies (coloured in green), with median values of $\lambda_{\rm{dark}} = 0.036 \pm 0.003$ and $\lambda_{\rm{bright}} = 0.026 \pm 0.002$, respectively. The significance in the difference between both distributions is confirmed by applying the Kolmogorov-Smirnov (KS) test, which results in a \textit{p}-value of $\sim10^{-5}$. 
Interestingly, the spin-parameter distribution of \emph{dark galaxies} matches the \cite{2001ApJ...555..240Bullock} model more closely than that of bright galaxies. This likely reflects the fact that these haloes have experienced little baryonic processing, preserving the ‘pristine’ angular momentum distribution predicted by \textit{N}-body simulations, whereas SF and feedback in bright galaxies can alter the spin of their host haloes. The larger spin parameters of \emph{dark galaxies} are directly related to their lower concentrations and may hinder the retention and condensation of gas to form stars, leading to lower gas densities as well (see also \citealt{Jimenez2020} and \citealt{2024ApJ...962..129Lee}).

\subsection{Gas content and thermodynamic properties}

We study the temperature and density of the gas in both dark and bright galaxies. The relation between these two properties at $z=0$ is shown in Fig.~\ref{fig: gas_tempdens} as a 2D histogram of gas particles aggregated over all galaxies in each sample. The top row corresponds to \emph{dark galaxies}, while the bottom row shows bright galaxies. Each column corresponds to a different simulation, from left to right {\hestia} 09\_18, 17\_11, 37\_11, and {\nivarialg}. Regardless of the particular simulation, \emph{dark galaxies} do not exhibit the characteristic star-forming regions in the $T$–$\rho$ phase diagram, as they lack gas at densities above the threshold required for SF, indicated by the vertical cyan and white line, for {\hestia} and {\nivarialg}, respectively. Bright galaxies contain on average a significant amount of dense gas, while \emph{dark galaxies} have little to no gas at  densities larger than $n_{\rm th}$. 

In {\hestia} simulations, bright galaxies display a region of dense gas in the upper-right part of the plots, owing to the implementation of an effective equation of state (EOS) that imposes an artificial pressure for gas above the SF density threshold (\citealt{2017MNRAS.467..179Grand}, following the \citealt{2003MNRAS.339..289Springel} model). This feature is instead absent in {\nivarialg}, where the gas is treated as an ideal gas at all densities and a polytropic effective EOS for star-forming gas is not imposed (e.g. \citealt{stinson2012,2015MNRAS.454...83Wang}). In {\nivarialg}, the more restrictive SF criterion ($n_{\rm th}\geq10\,\rm cm^{-3}$) makes it more difficult for galaxies to convert their gas into stars: as a consequence, a given \emph{dark galaxy} might be able to form some stars in {\hestia}, but not in {\nivarialg}. It shall be noted that the absence of gas particles above the SF density threshold in the $T$–$\rho$ diagram for {\nivarialg} bright galaxies, simply indicates that these systems are not currently forming stars at $z=0$. However, they did experience SF at earlier epochs, in contrast to bright galaxies in {\hestia}, which are predominantly star-forming at $z=0$\footnote{We note that, in the current {\nivarialg} sample, all bright galaxies appear quenched at $z = 0$. This may reflect environmental effects within the Local Group volume (e.g. cosmic-web stripping or backsplash objects), and will be explored in future higher-resolution {\nivarialg} simulations.}.

    \begin{figure*}
    \centering
    \includegraphics[scale=0.195, trim={0cm 1cm 1cm 0cm}, clip]{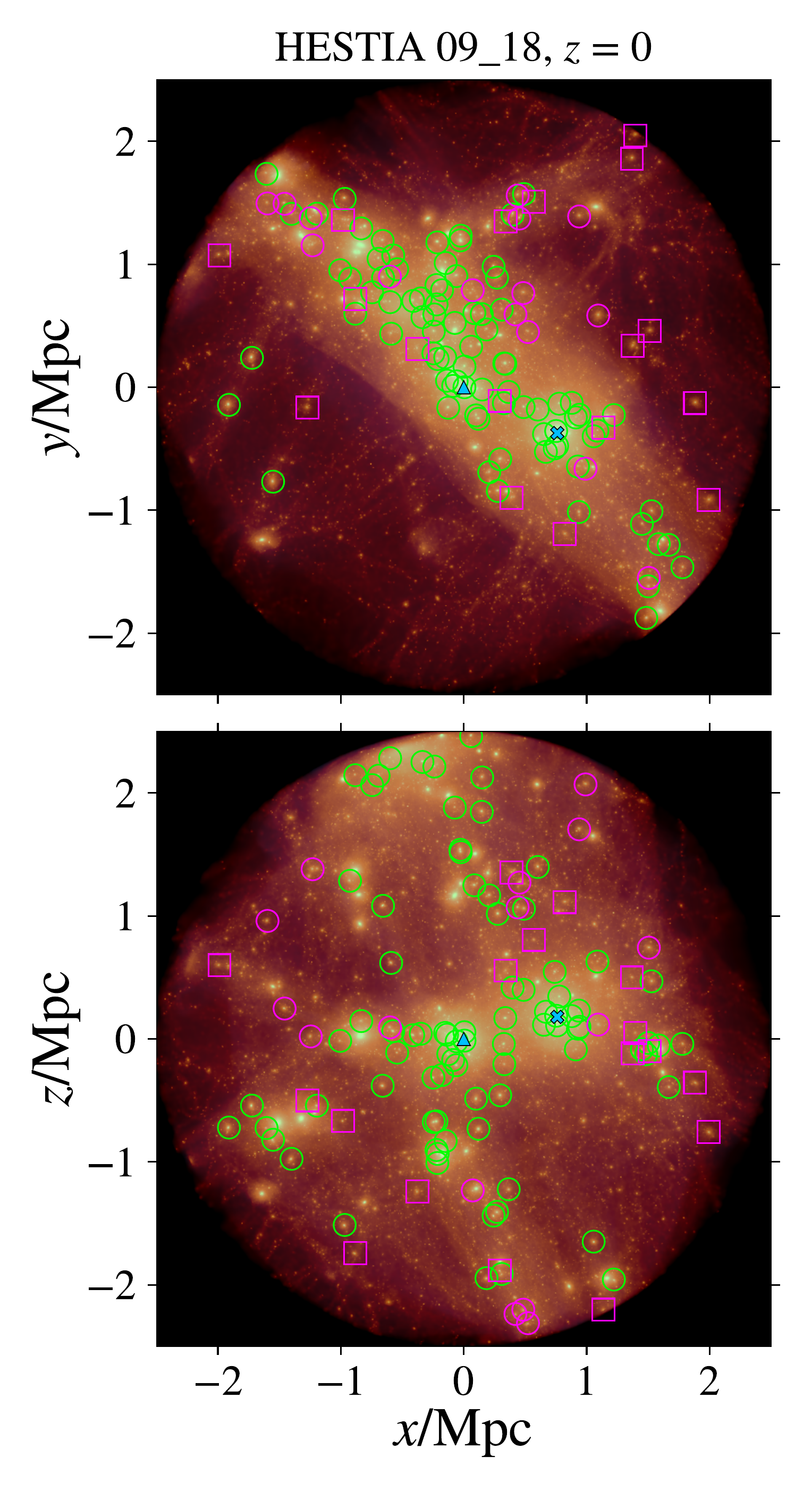} 
    \includegraphics[scale=0.195, trim={4.4cm 1cm 1cm 0cm}, clip]{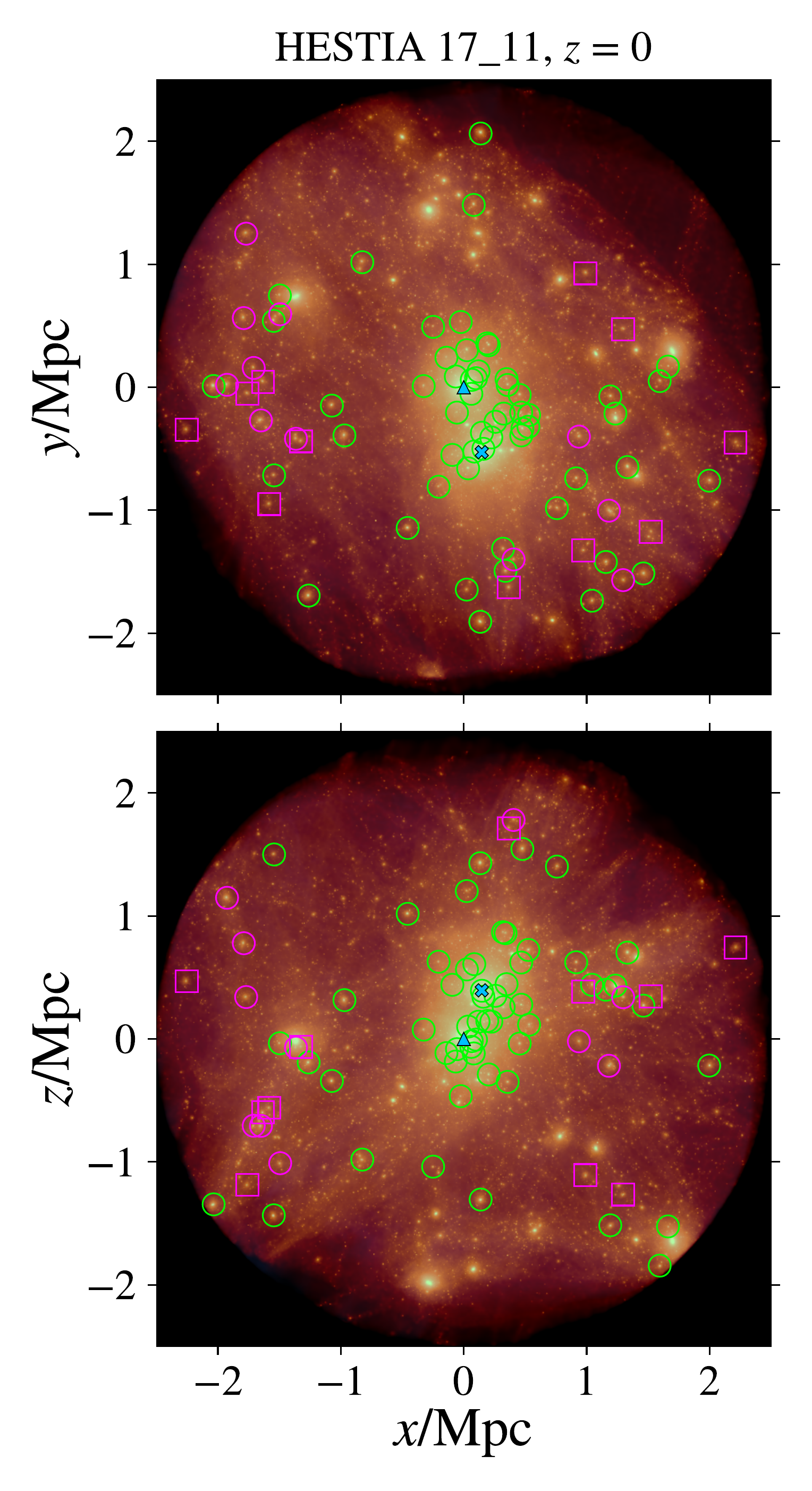} 
    \includegraphics[scale=0.195, trim={4.4cm 1cm 1cm 0cm}, clip]{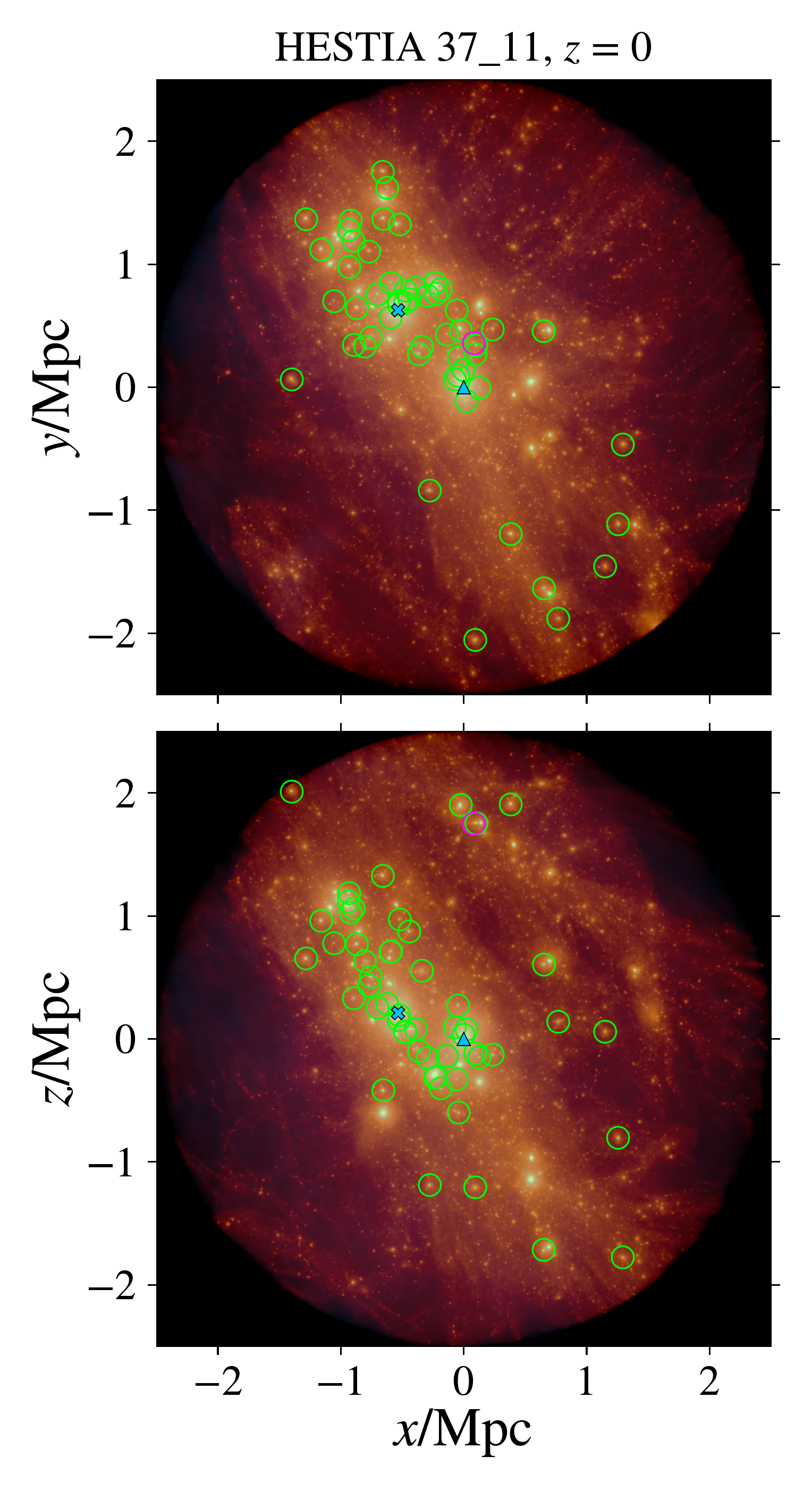} 
    \includegraphics[scale=0.195, trim={0.7cm 1cm 1cm 0cm}, clip]{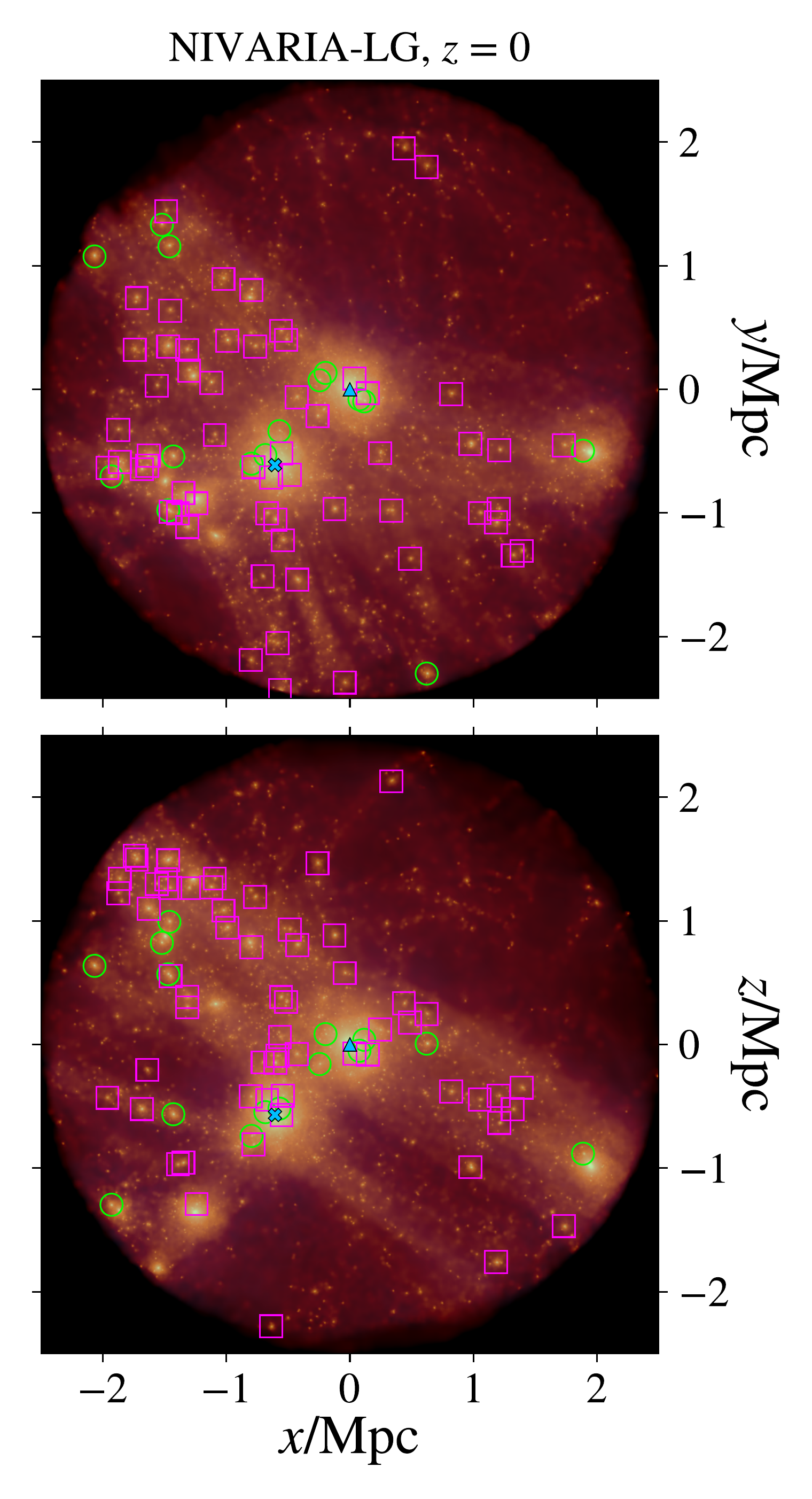}
   \caption{Projected positions of dark (magenta) and bright (green) galaxies in the $x-y$ (\textit{top row}) and $x-z$ (\textit{bottom row}) planes at $z = 0$, within a sphere of 2.5 Mpc of radius from the MW. From the \textit{left} to \textit{right}, we present the three {\hestia} simulations (09\_18, 17\_11, and 37\_11) and {\nivarialg}. Circles indicate \emph{dark galaxies} with stars, whereas squares indicate the starless \emph{dark galaxies}. The background shows the density of DM in orange and the gas density in yellow. The region is centred on the MW, shown as a blue triangle, while  M31 is represented as a blue cross mark.} 
          \label{fig: spatialprojections}%
    \end{figure*}

    \begin{figure}
    \centering
    \includegraphics[width=8cm]{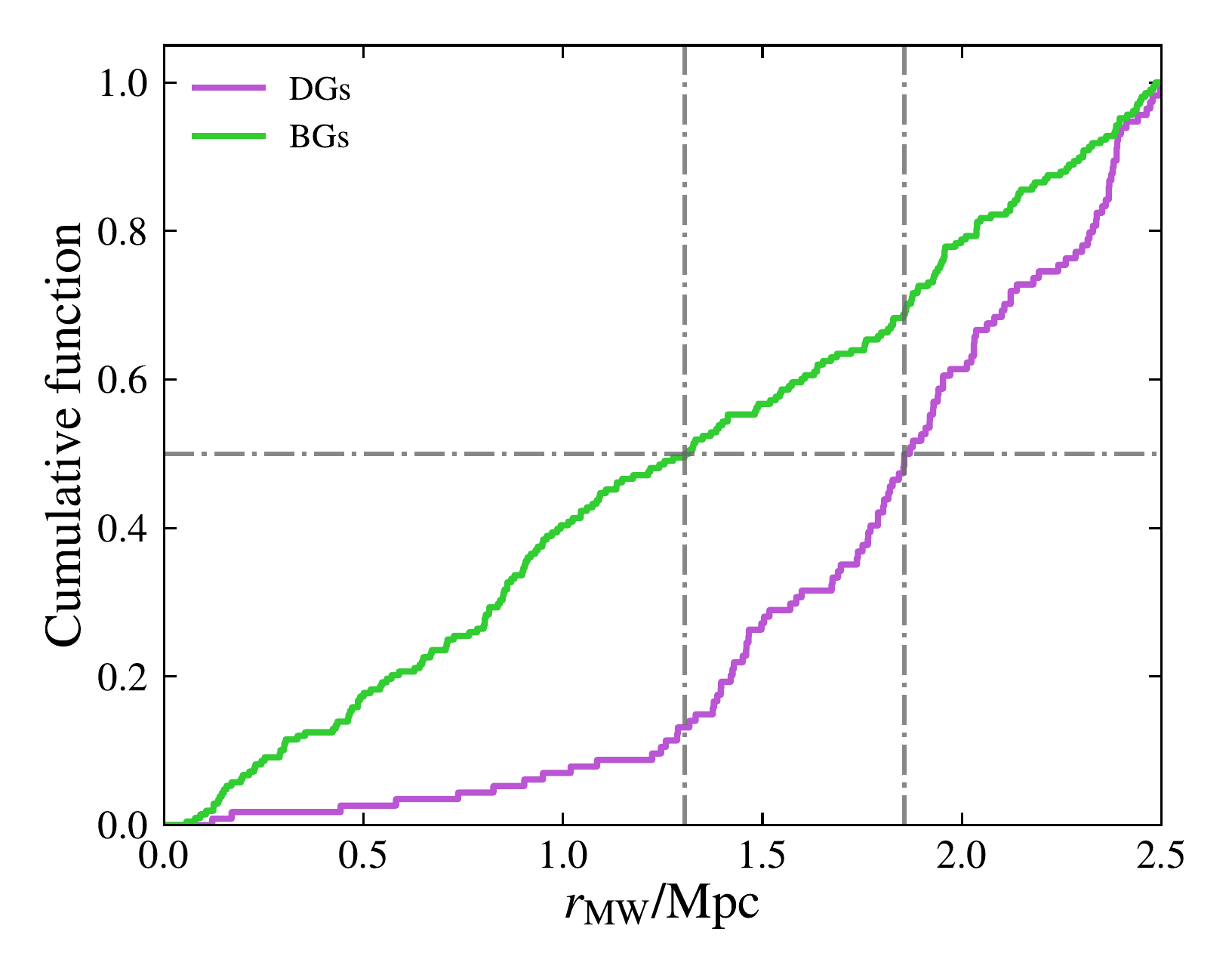}
   \caption{Cumulative distribution of the radial distances of dark (magenta) and bright (green) galaxies with respect to the MW analogues for all four simulations at $z = 0$. The horizontal grey line marks the mean of the distributions, while the two vertical lines indicate the distances within which 50$\%$ of dark and bright galaxies are found, corresponding to $r_{\rm MW} = 1.9 \rm\ Mpc$ and $r_{\rm MW} = 1.3 \rm\ Mpc$, respectively.} 
          \label{fig: distanceMW}%
    \end{figure}    

As demonstrated in \cite{2017MNRAS.465.3913Benitez_Llambay}, gas within \emph{dark galaxies} should exhibit a characteristic temperature–density relation arising from the combined effects of gas cooling at high densities and photoheating at low densities. This relation features a peak at $T\sim 4\times 10^4\,\mathrm{K}$ and $\rho \sim 10^{-4.8}\,\mathrm{cm}^{-3}$. This prediction, shown in the top, \emph{dark galaxies} panels of  Fig.~\ref{fig: gas_tempdens} as a solid red line, is  in very good agreement with  results from our simulations. This relation reflects an approximate hydrostatic equilibrium between the gas and the halo gravitational potential, as well as thermal equilibrium with the UV background under the assumption of spherical symmetry. The gas properties of \emph{dark galaxies} thus establish the conditions for little to no SF and are consistent with the characteristics of their underlying DM haloes. Lower halo concentrations, lower halo masses at the epoch of reionisation, and higher spin parameters compared to bright galaxies of similar mass, promote under-dense gas to remain in hydrostatic and thermal equilibrium within its DM matter halo and to never form stars.

\subsection{Spatial distribution and environment} \label{subsect: environment}
In order to assess the impact of the intergalactic medium on the properties of \emph{dark galaxies}, we analyse in this section whether dark and bright galaxies inhabit different environments and how the density of these surroundings evolves over time.
Figure~\ref{fig: spatialprojections} shows the distribution of dark (magenta) and bright (green) galaxies in both the $x-y$ and $x-z$ projections for the four simulation runs at the present epoch. The figure displays a region centred on the MW analogue and extending to a radius of $2.5 \rm\ Mpc$. The background represents the superimposed DM and gas density fields. Bright galaxies are generally more concentrated in high-density regions, closer to the two main central galaxies. In contrast, the distribution of \emph{dark galaxies} appear to be more spatially extended, occupying larger radial distances, and residing predominantly farther from the central regions and the bulk of the main filament. This trend is less evident in {\nivarialg}, because the sample of bright galaxies is significantly smaller than that of \emph{dark galaxies}; nevertheless, the trend is still present.

    \begin{figure*}
    \centering
    \includegraphics[width=9cm]{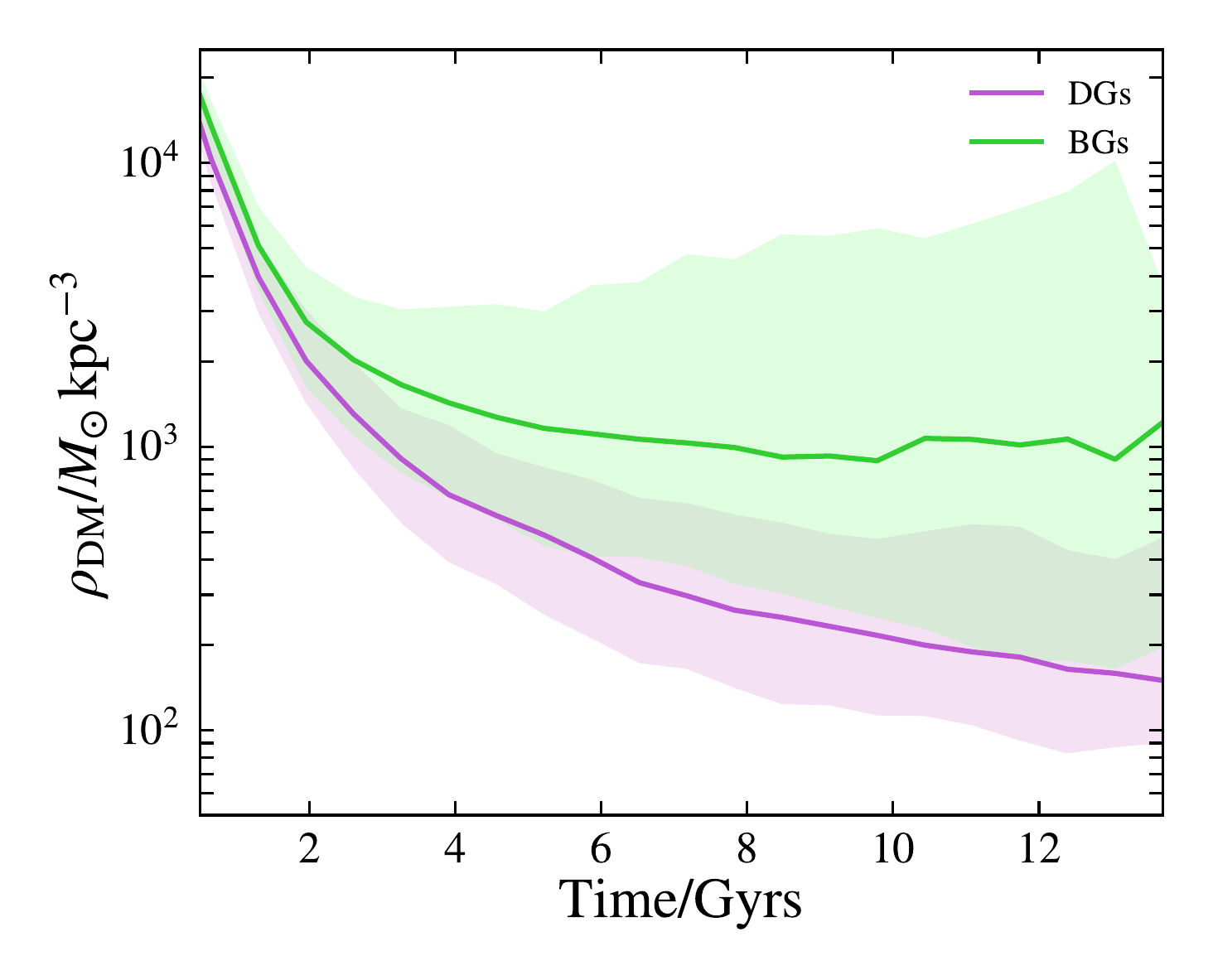}
    \includegraphics[width=9cm]{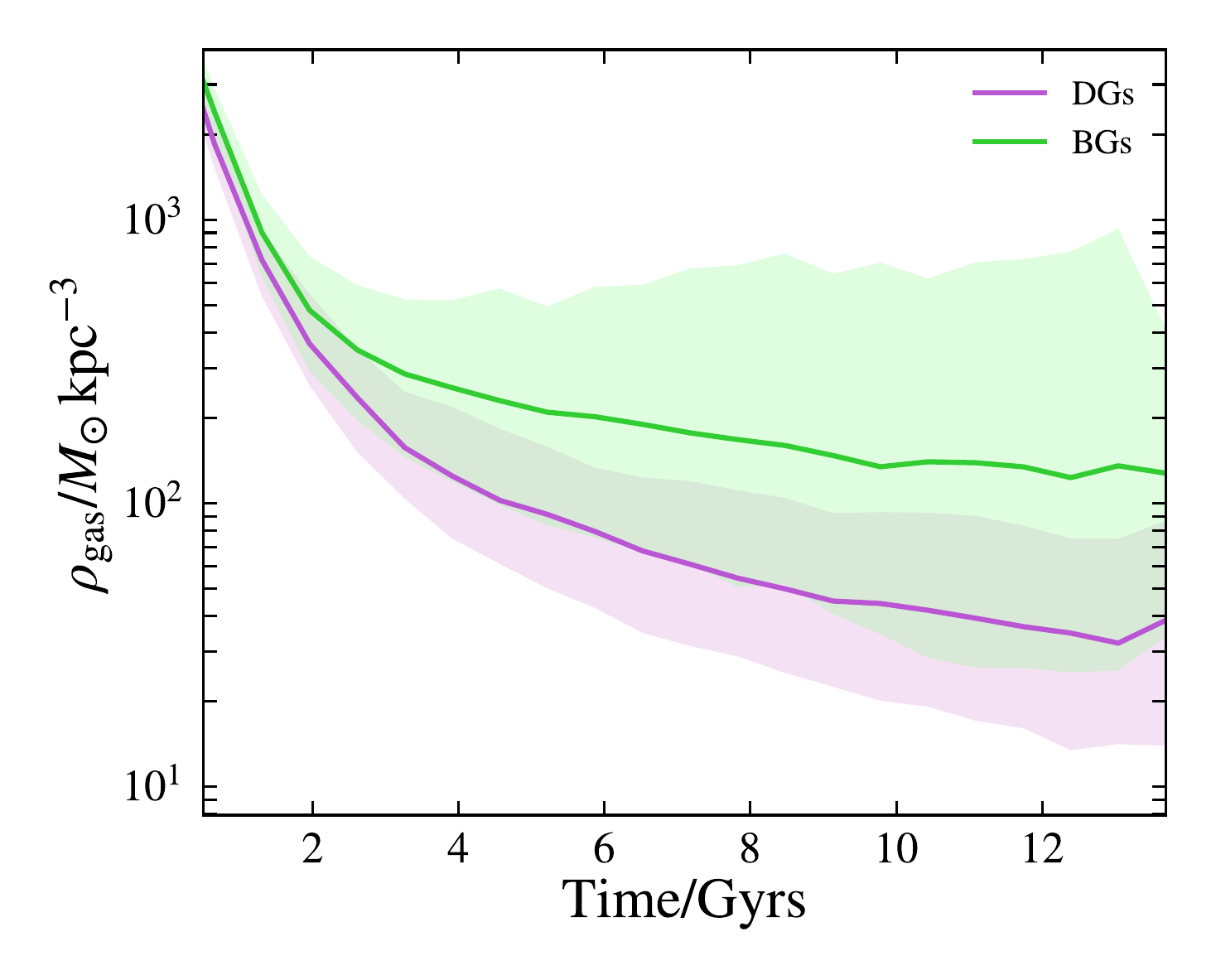}
   \caption{Evolution of median DM (\textit{left panel}) and gas (\textit{right panel}) density of the environment around the full sample of \emph{dark galaxies} in magenta and bright galaxies in green, combining all four simulations. The shaded backgrounds are the regions limited by the $16^{\rm th}$ and $84^{\rm th}$ percentiles of the distributions. The environment is defined as a spherical shell between 1 $R_{\rm vir}$ and $7\,R_{\rm vir}$ of each halo, at each redshift.}
          \label{fig: dens_envirn_joinmedian}%
    \end{figure*}

The larger distances from the MW at which \emph{dark galaxies} are found are further illustrated in Fig.~\ref{fig: distanceMW}, which shows the cumulative distribution of the radial distances of all dark (magenta) and bright (green) galaxies from the MW analogues. The difference between the two radial distributions is again confirmed by a KS test, yielding a \textit{p}-value of $\sim10^{-10}$. This low value indicates that the two samples are statistically distinct. We find that 50$\%$ of the \emph{dark galaxy} sample lies within $1.9 \rm\ Mpc$ of the MW, whereas 50$\%$ of bright galaxies are located within $1.3 \rm\ Mpc$. Consequently, we expect a large fraction of \emph{dark galaxies} to be found in the outskirts of the Local Group.

We take a step further and examine the evolution of the DM and gas densities within a spherical shell extending from $1\, R_{\rm\ vir}$ to $7\, R_{\rm\ vir}$ around each dark and bright galaxy identified at different redshifts. The results are shown in Fig.~\ref{fig: dens_envirn_joinmedian}, where the median densities of all dark and bright galaxies from the four simulations are displayed in magenta and green, respectively. The shaded background regions represent the $16^{\rm th}$ and $84^{\rm th}$ percentiles of each distribution. Consistent with the results shown in Fig.~\ref{fig: spatialprojections}, bright galaxies reside in environments with higher DM and gas densities than \emph{dark galaxies}. These results indicate that \emph{dark galaxies} originate in lower-density regions within the Local Group and remain in such environments throughout their lifetime. Their persistently lower DM and gas density environment make \emph{dark galaxies} less susceptible to galaxy interactions, mergers, and gas inflows, allowing their gas to remain diffuse and in equilibrium within their DM halo rather than collapsing. 
These results are in agreement with previous work from \citet{2017MNRAS.465.3913Benitez_Llambay}, who showed that gas-rich RELHICs
inhabit the under-dense outskirts of the {\apostle} simulations.
In Appendix~\ref{appendix_1}, we present the same density evolution as shown in Fig.~\ref{fig: dens_envirn_joinmedian}, but separately for each simulation. We also include the evolution of the median $R_{\rm vir}$ of all dark and bright galaxies in the upper panels of Fig.~\ref{fig: dens_envirn_median}, as a reference to the physical scales involved with respect to the defined environment.

\subsection{{\HIsc} detectability with FAST}\label{subsect: HIFAST}
Objects such as \emph{dark galaxies}, which exhibit little to no luminous matter, are inherently difficult to identify observationally. Nevertheless, if they contain a substantial reservoir of 
{\HIsc} they may be detectable as gas clouds via their $21\,\rm cm$ line emission, using facilities such as the FAST radio telescope. In this section, we present estimates of the detectability of \emph{dark galaxies} based on their {\HIsc} content, selecting FAST as the instrument of choice due to its large  sky coverage and high sensitivity. 

We estimate the amount of neutral atomic hydrogen in the \texttt{HESTIA} sample by adopting the empirical approach presented in \cite{2024A&A...690A.286Arjona}. This method builds upon the phenomenological prescription introduced by \cite{2017MNRAS.466.3859Marinacci}, itself based on \citet{2008AJ....136.2782Leroy}. In this framework, the ratio between the molecular and atomic hydrogen column densities is fitted with a functional form that depends on the gas mid-plane pressure, which then allows us to infer the {\HIsc} atomic fraction.
On the other hand, the {\HIsc} fraction in {\nivarialg} is measured following the self-shielding approximation detailed in \cite{2013MNRAS.430.2427Rahmati}, which uses the radiative transfer model from \cite{2008MNRAS.389..651Pawlik, 2011MNRAS.412.1943Pawlik}. 

Figure~\ref{fig: mhalo_mHI} shows the {\HIsc} masses of the \emph{dark galaxy} sample in {\hestia} simulations (blue shades) and {\nivarialg} (red), as a function of their DM halo masses. We include Cloud-9 as a reference\footnote{Note here Cloud-9 is not within the range of distances of our analysis, i.e. $d_{\rm{Cloud-9}} \sim 4.7\,\rm Mpc$.}. We can observe a very strong   correlation between the halo mass and the {\HIsc} gas in each \emph{dark galaxy}. We perform an exercise to furnish a quick estimate on the detectability of these \emph{dark galaxies} with FAST. Details on how the FAST limits are calculated can be found in Appendix~\ref{appendix_2}. The two shaded areas in Fig.~\ref{fig: mhalo_mHI} represent the region of detectability  within $1\,\rm Mpc$ and $2.5\,\rm Mpc$, according to the values obtained for the minimum {\HIsc} mass detectable with FAST. Here, it can be seen that up to 18 of our \emph{dark galaxies} fall in the region corresponding to distances within $1\,\rm Mpc$, and up to 10 of those fall in the region within $2.5\,\rm Mpc$. These maximum values are set by the {\nivarialg} simulation, which contributes with the largest number of \emph{dark galaxies} within the detectability region compared to the rest of simulations. It is also clear that several \emph{dark galaxies} are below the detection limits for such distances. 

In Fig.~\ref{fig: Cloud9_mHI_rMW} we further clarify the distance dependence by showing  {\HIsc} masses of  \emph{dark galaxies}  with respect to their distance from the simulated MW analogues. The grey dotted-dashed line represents the {\HIsc} mass limit with respect to the distance, as given by Eq.~\ref{eq: HImass}. All \emph{dark galaxies} located above this line should be detectable by FAST in terms of their total {\HIsc} mass and distance. The area of detectability is shown as the grey shaded region. In this case, the number of \emph{dark galaxies} falling within said region is between 0 and 18, according to each one of the simulation runs used (where {\nivarialg} provides the largest estimate, and {\hestia} 37\_11 the lowest).

We further explore the detectability of these galaxies by analysing their {\HIsc} column density profiles, which are shown in Fig.~\ref{fig: columnHIdensity}, for {\hestia} in the left panel and for {\nivarialg} in the right panel. The profiles are coloured according to the {\HIsc} mass of each \emph{dark galaxy}, where darker colours indicate larger {\HIsc} masses. The coloured vertical dotted-dashed lines indicate the softening length of each simulation, below which profiles are not well resolved. Since FAST can reach down to a median 3$\sigma$ column density of $N_{\rm{H}\,{\textsc i}}=2 \times10^{17} \rm{cm^{-2}}$ (\citealt{2024MNRAS.534..202Pan}), we show this limit in the panels as a grey horizontal dotted-dashed line. Hence, all galaxies whose profiles lie above this threshold and within the grey shaded area would be detectable by FAST. We can see that all profiles with values exceeding the FAST column density detection limit (and beyond the softening length) have {\HIsc} masses $> 10^{4}\,\rm \textit{M}_{\odot}$. These results are added to Fig.~\ref{fig: mhalo_mHI} and Fig.~\ref{fig: Cloud9_mHI_rMW}, where we now indicate the galaxies corresponding to these minimum detectable column densities using concentric circles. Therefore, there is a trend for which the most massive \emph{dark galaxies} in DM are generally those that can be detectable, in agreement with  results from \cite{2017MNRAS.465.3913Benitez_Llambay}. On the other hand, since the simulated \emph{dark galaxies} have been centred and set in a face-on orientation, the $N_{\rm{H}\,{\textsc i}}$ values shown in Fig.~\ref{fig: columnHIdensity} are actually a lower limit, and could be slightly larger when placing the galaxies edge-on. 

    \begin{figure}
    \centering
    \includegraphics[width=9cm]{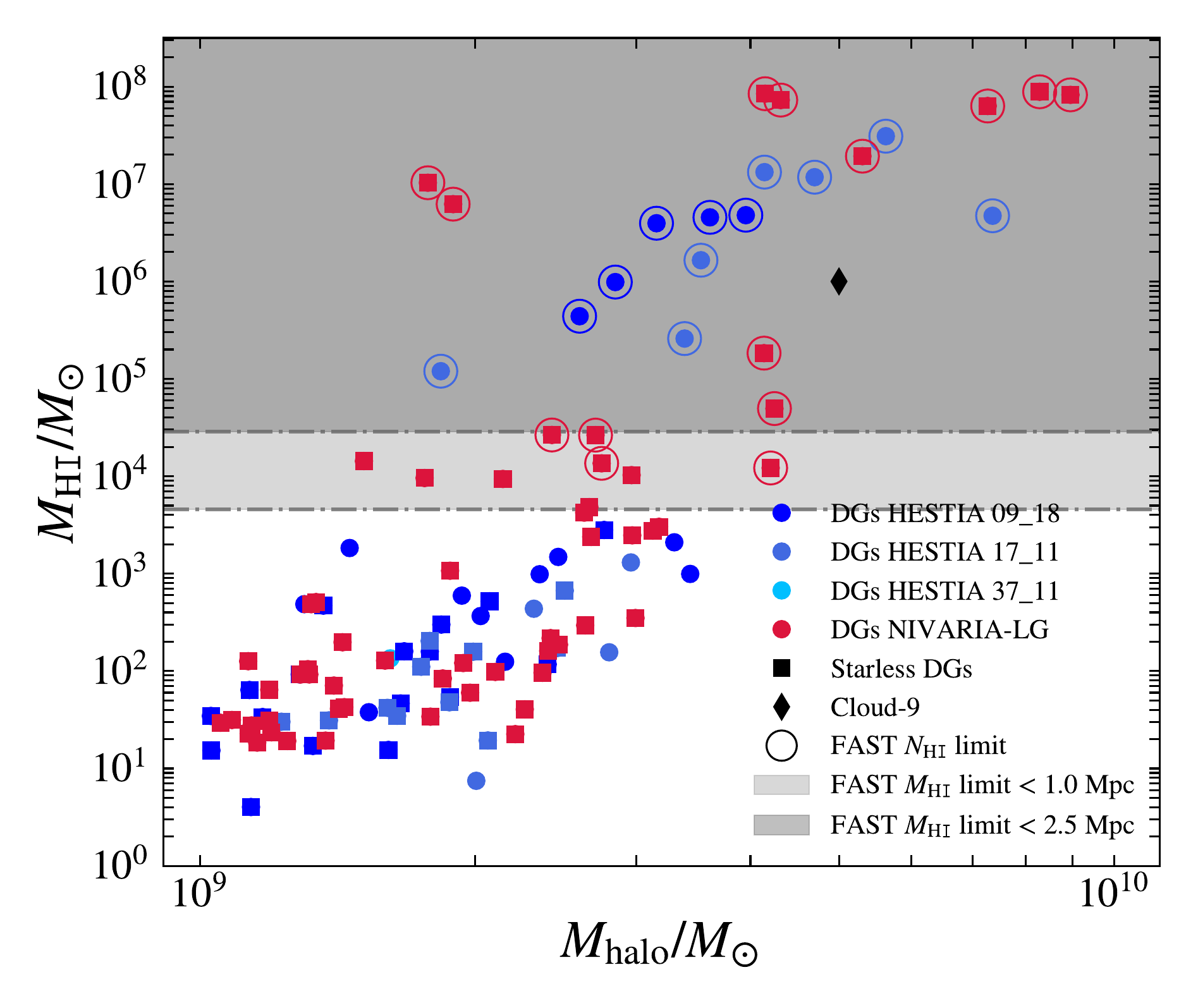}
    \caption{Mass of {\HIsc} versus DM halo mass for \emph{dark galaxies} in each simulation, with {\hestia} indicated as blue shades, while {\nivarialg} is shown in red. Squares symbols denote  starless \emph{dark galaxies}. The grey shaded regions represent galaxies with {\HIsc} masses that could be detectable with FAST in 1 hour of integration time and at distances of $< 1\,\rm Mpc$ and $< 2.5\,\rm Mpc$ from the MW analogues. The concentric circles indicate  \emph{dark galaxies} whose central {\HIsc} column densities exceed  the FAST detection limit,  $N_{\rm{H}\,{\textsc i}}$. We include Cloud-9 as a black diamond.}
          \label{fig: mhalo_mHI}%
    \end{figure}

    \begin{figure}
    \centering
    \includegraphics[width=9cm]{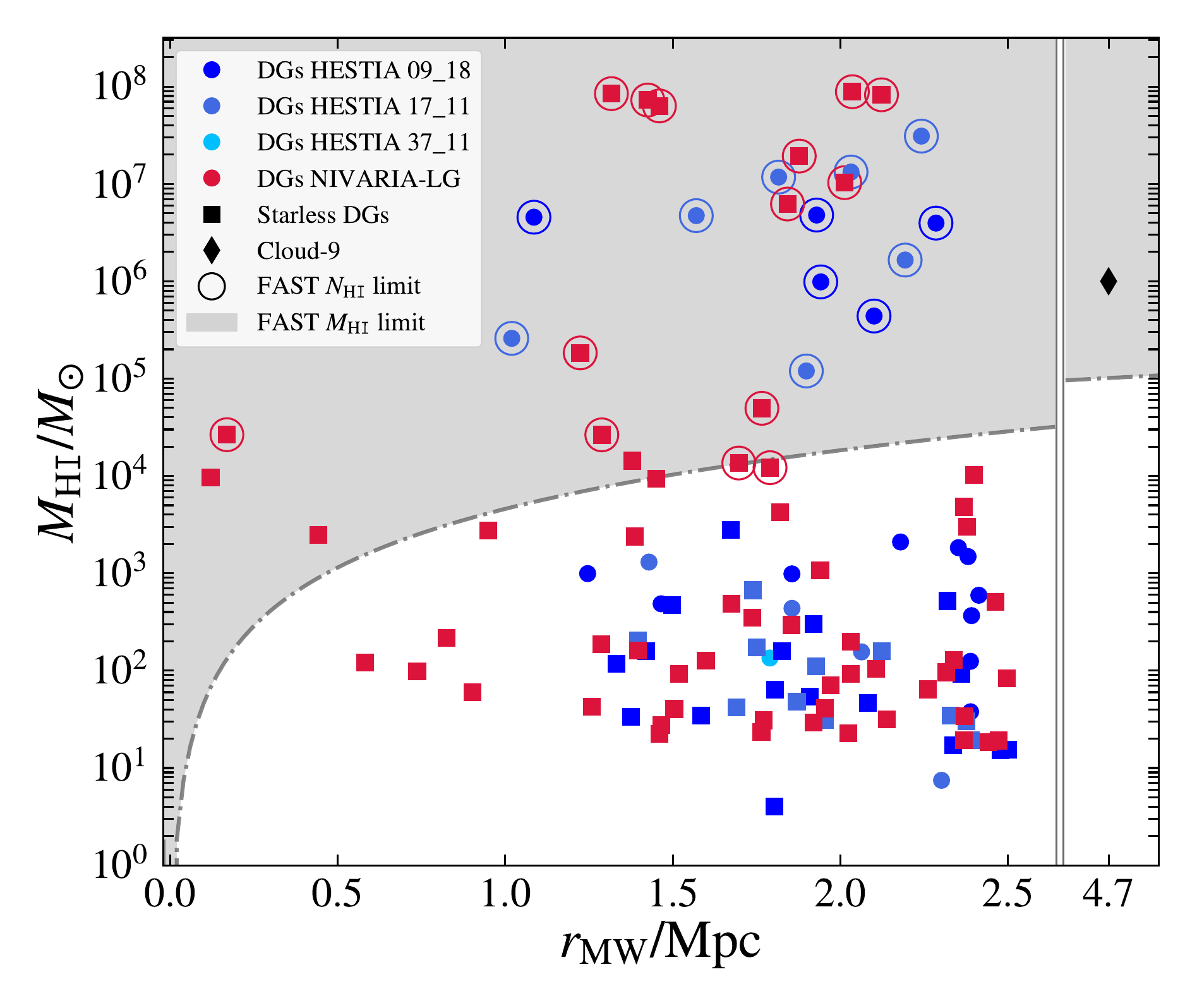}
    \caption{{\HIsc} mass versus radial distance of \emph{dark galaxies} with respect to the MW analogues in each simulated run, with {\hestia} indicated as blue shaded symbols and {\nivarialg} as red symbols. Squares correspond to starless \emph{dark galaxies}. The grey dotted-dashed line represents the minimum {\HIsc} mass that galaxies must have to be detectable by FAST, as a function of their distance from us. \emph{Dark galaxies} likely to be detectable by FAST are those that fall within the shaded region. The concentric circles indicate  \emph{dark galaxies} whose central {\HIsc} column densities  $N_{\rm{H}\,{\textsc i}}$ exceed the FAST detection limit. Cloud-9 is included as a black diamond at $d = 4.7\,\rm Mpc$.} 
          \label{fig: Cloud9_mHI_rMW}%
    \end{figure}

    \begin{figure*}
    \centering
    \includegraphics[scale=0.36, trim={0cm 0cm 0.95cm 0cm}, clip]{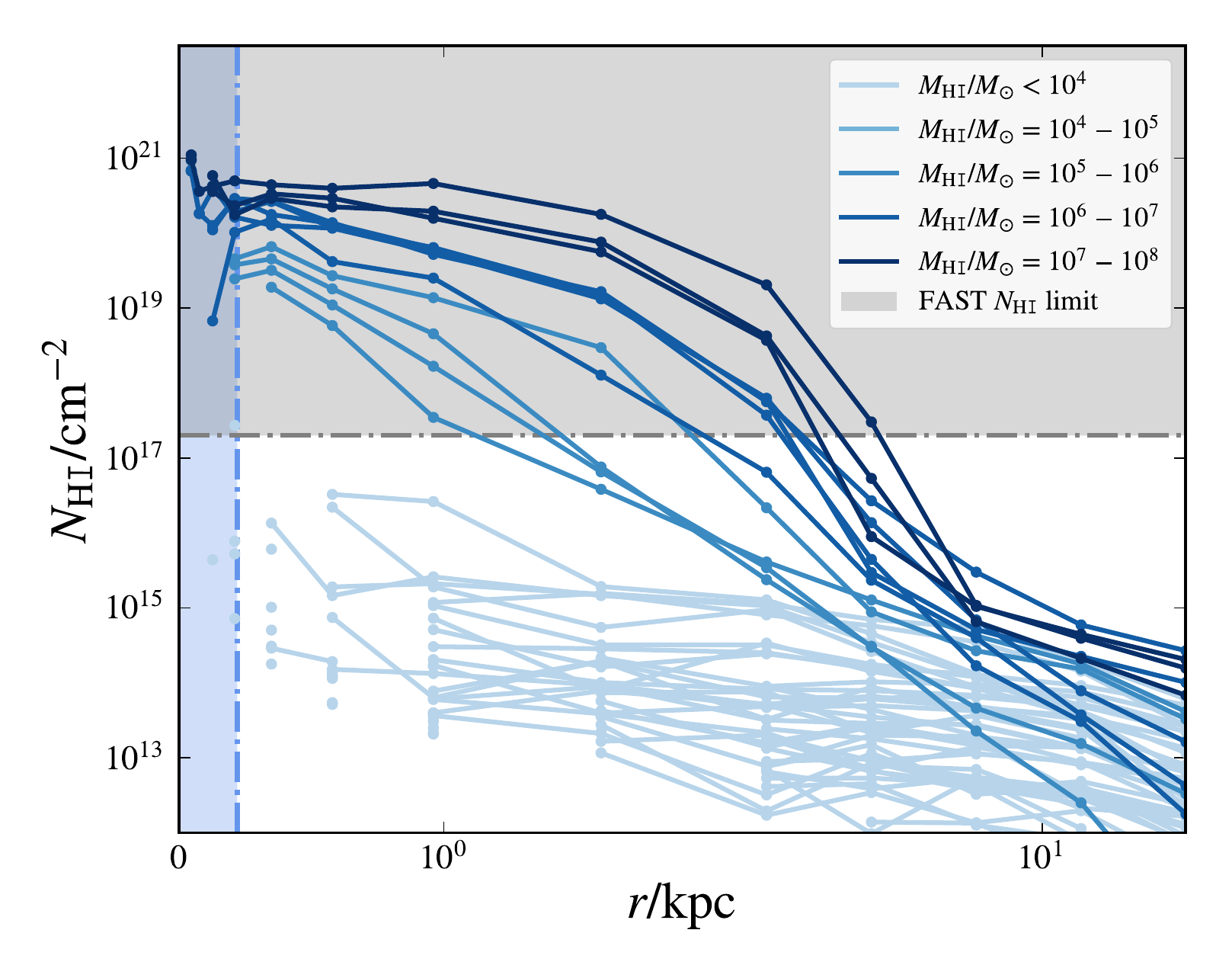}
    \includegraphics[scale=0.36, trim={3.8cm 0cm 0cm 0cm}, clip]{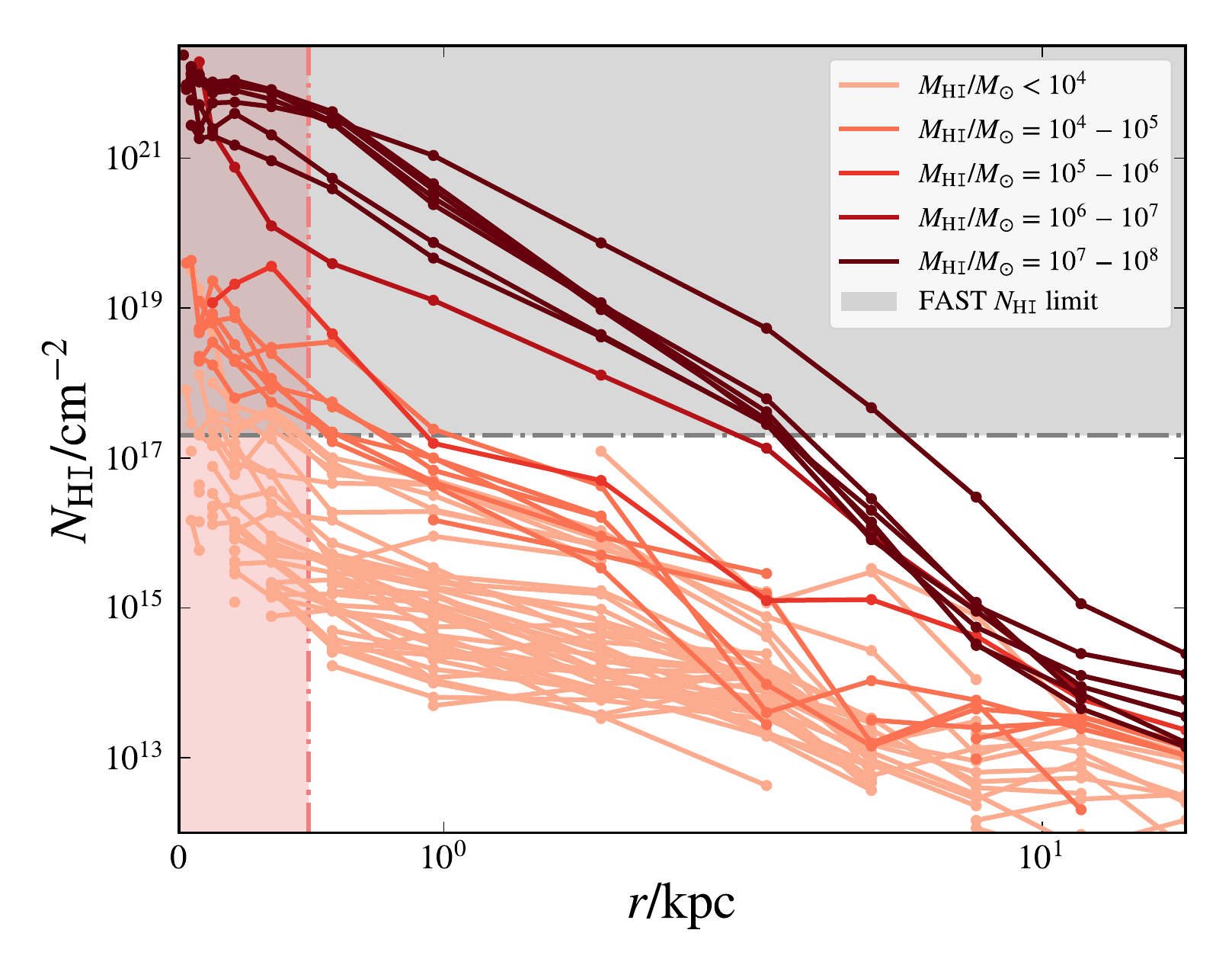}
   \caption{{\HIsc} column density profiles of \emph{dark galaxies} for the {\hestia} simulations (\textit{left}) and the {\nivarialg} simulations (\textit{right}), coloured by {\HIsc} mass. The grey dotted-dashed line represents the minimum {\HIsc} column density value that FAST can reach, which is $N_{\rm{H}\,{\textsc i}}=2 \times10^{17}\,\rm{cm^{-2}}$. The coloured shaded area in each panel shows the softening length corresponding to each simulation. A smaller number of \emph{dark galaxies} could be detected by the less sensitive ALFALFA Survey, whose sensitivity is $N_{\rm{{\HI}}} \gtrsim 10^{18}\, \rm{cm}^{-2}$.}
          \label{fig: columnHIdensity}%
    \end{figure*} 

That said, the total number of \emph{dark galaxies} with both sufficient H \textsc{i} mass and column densities to be detectable by FAST is up to 14 when using the NIVARIA-LG simulation, which therefore provides an upper limit: these galaxies are indicated as red symbols with concentric circles within the shaded grey region in Fig.~\ref{fig: Cloud9_mHI_rMW}. Conversely, only 5 and 7 \emph{dark galaxies} would be detectable by FAST, according to the {\hestia} simulations (seeds 09\_18 and 17\_11). Since the simulations consider the whole sky, these values are effectively smaller when we account for the fact that FAST is located in the northern hemisphere and covers only $\sim58\%$ of the total sky (e.g., \citealt{2020Innov...100053Qian}). 

Consequently, the upper limit on the number of \emph{dark galaxies} detectable within 2.5 Mpc from the MW,  taking into account the FAST sky coverage and minimum {\HIsc} mass and column density, is 8. This is our preferred value, based on the more realistic {\nivarialg} simulations which, notably, produce all starless \emph{dark galaxies}.

\section{Discussion}\label{sect: discussion}

We find that bright galaxies in our sample include a substantially higher fraction of satellites than \emph{dark galaxies}, indicating that luminous systems are more frequently bound to larger hosts, while \emph{dark galaxies} preferentially reside in isolation. This environmental distinction also appears in the satellite statistics of \cite{2024ApJ...962..129Lee}, where star-rich galaxies exhibit higher satellite fractions than star-poor or starless systems. In our simulations, these differences naturally place bright and \emph{dark galaxies} in contrasting environments: bright galaxies experience more mergers, interactions, and tidal capture, whereas \emph{dark galaxies} remain largely detached from dense environments and massive hosts. This is fully consistent with the environmental-density trends presented in Sec.~\ref{subsect: environment}, which show that \emph{dark galaxies} lie farther from the main filaments and central regions of the Local Group. Moreover, as shown in Fig.~\ref{fig: dens_envirn_joinmedian}, not only do \emph{dark galaxies} inhabit lower-density regions today, but they also originate in underdense environments, which can directly suppress SF by limiting gas accretion. Although at very high redshift this scenario holds uncertainties
due to the much higher and uniform global density of the
Universe and the limitations of the simulations, the
median trends in Fig.~\ref{fig: dens_envirn_joinmedian} still show a difference of the order of $\sim0.1$ dex in both DM and gas density at early epochs. This picture is also supported by observational results from \cite{2025ApJS..279...38Kwon}, who find that potential \emph{dark galaxy} candidates in the ALFALFA survey preferentially occupy low-density environments.

The distinct formation histories of bright and \emph{dark galaxies} further reinforce this scenario. As shown in Fig.~\ref{fig: masses_z}, \emph{dark galaxies} assemble later ($z_{\rm dark}=1.7-2.0$) than bright galaxies ($z_{\rm bright}=2.6-3.2$), resulting in systematically lower concentrations (Fig.~\ref{fig: mhalo_cNFW}). Combined with their higher spin parameters, this implies that \emph{dark galaxies} inhabit more extended, less concentrated haloes that retain under-dense gas, below the SF threshold. 

These findings are in excellent agreement with results from \cite{2017MNRAS.465.3913Benitez_Llambay} and \cite{2024ApJ...962..129Lee}, even though their studies employed a different simulation suite, numerical methods, resolution, and initial conditions.
An important difference with the work of \cite{2024ApJ...962..129Lee}, however, is that they do not require their sample of \emph{dark galaxies} to necessarily contain {\HIsc} gas, while in our selection we explicitly require {\HIsc} gas to be present, so as to assess the  detectability of \emph{dark galaxies} with observational surveys. The comparison with \cite{2024ApJ...962..129Lee} should therefore be considered qualitative, since the \emph{dark galaxy} samples in our work and theirs are selected using different underlying criteria.

In a recently submitted work, \cite{2025arXiv251116726Zheng} studied a sample of \emph{dark galaxies} in the {\hestia} and {\auriga} simulations, focusing on faint systems with $M_{g} > -10$ and $M_{\rm{H}\,{\textsc i}} > 10^{5}\, M_{\odot}$. In their sample, only one completely starless object is identified, found in {\auriga}. As shown in our Fig.~\ref{fig: mhalo_mHI}, we likewise do not find any starless \emph{dark galaxies} in {\hestia} with $M_{\rm{H}\,{\textsc i}} > 10^{5}\, M_{\odot}$, although such objects do appear in {\nivarialg}, which we attribute to the more restrictive SF conditions adopted in this simulation.
Despite the different selection criteria, our results are broadly consistent with those of \cite{2025arXiv251116726Zheng}. Their study primarily examines simulations that employ similar subgrid physics ({\hestia} and {\auriga}), whereas we compare dark and luminous galaxies across two Local Group simulations that adopt different physical models and SF prescriptions, allowing us to assess the robustness of \emph{dark galaxy} predictions across distinct simulation setups. \cite{2025arXiv251116726Zheng} also provide predictions for the number density of \emph{dark galaxies} within a given volume. However, differences in selection criteria make direct comparisons non-trivial.

\section{Conclusions}\label{sect: conclusions}
In this work, we analysed the three high-resolution \hestia\ simulation runs \citep{2020MNRAS.498.2968Libeskind} and the new \nivarialg\ simulation \citep{ContrerasSantos_etal} to derive the properties of \HIsc-rich \emph{dark galaxies} within Local Group–like volumes simulated with the same constrained initial conditions but with different numerical codes and galaxy-formation models ({\arepo} for the \hestia\ simulation, {\gasoline} for \nivarialg). Across the four runs, we identified a total of 114 \emph{dark galaxies} in the DM halo mass range $10^9-10^{10}\,{M_\odot}$, of which only six are satellites of larger galactic systems at $z=0$. We then compared their properties to those of a control sample of bright galaxies within the same halo mass range. Our main conclusions are summarised below.\\\\

 \emph{Dark galaxies}: 
   \begin{itemize}
     \item  exist in Local Group simulations run with different codes and feedback models, with their abundance depending on the mass of the Local Group itself and, critically, on the adopted SF density threshold $n_{\rm th}$. Simulations that employ a more realistic, higher $n_{\rm th}$, produce significantly more \emph{dark galaxies} than those that do not (up to $\sim60$ in the {\nivarialg} run, compared to 17, 11, and 0 starless \emph{dark galaxies} arising, respectively, in the three {\hestia} runs, as shown in Table~\ref{tab: sample});
    \item contain a similar total amount of gas as their bright-galaxy counterparts, although only the most massive \emph{dark galaxies}, with $M_{\rm halo} > 10^{9.5}\,M_\odot$, have large {\HIsc} gas reservoirs comparable to those found in bright galaxies, i.e. $M_{\rm{H}\,{\textsc i}} > 10^{5}\,M_\odot$ (Fig.~\ref{fig: mhalo_mstar_mgas}). Such gas  cannot form stars efficiently because of its low density, being in hydrostatic equilibrium within the gravitational potential of its halo (Fig.~\ref{fig: gas_tempdens});
    \item have lower masses than bright galaxies at the epoch of reionisation, and throughout cosmic time (Fig.~\ref{fig: masses_z}), which render them dark, since galaxy formation can occur only in haloes whose masses exceed a critical threshold \citep{Benitez-Llambay2020};
    \item have $z=0$ halo masses peaking at $\sim2\times10^9\,M_{\odot}$    (Fig.~\ref{fig: mhalo_hist}), higher spin parameters (Fig.~\ref{fig: spinparameter}), and lower concentrations (Fig.~\ref{fig: mhalo_cNFW}) than bright galaxies, consistent with their late formation times (Fig.~\ref{fig: masses_z});
    \item are located farther from the MW than luminous galaxies (Fig.~\ref{fig: distanceMW}), with about 80$\%$ residing at distances greater than $1.5\,\rm Mpc$, and inhabit lower-density regions of the Local Group (Fig.~\ref{fig: spatialprojections});
    \item they formed and evolved across redshift within regions characterised by persistently lower DM and gas densities compared to the environments of bright galaxies, making them less susceptible to galaxy interactions, mergers, and gas inflows (Fig.~\ref{fig: dens_envirn_joinmedian});
    \item can be detected via their {\HIsc} emission.
    Using the sky coverage of the new FAST radio telescope, along with its minimum detectable {\HIsc} mass and column-density sensitivity, we predict that up to 8 \emph{dark galaxies} should be detectable in {\HIsc} within $2.5\,\rm Mpc$ of the Local Group (Figs.~\ref{fig: mhalo_mHI} and~\ref{fig: Cloud9_mHI_rMW}).
   \end{itemize}

The {\nivarialg} simulation adopts a SF density threshold, $n_{\rm th}=10\,\rm{cm^{-3}}$, that is more realistic than the {\hestia} one, reflecting the densities of clumps within typical giant molecular clouds ($10-100\,\rm cm^{-3}$). It  produces a great match of the $M_{\star}$–$M_{\rm halo}$ relation over several orders of magnitude in halo mass (Fig.~\ref{fig: allmhalo_mstar}) and is the only simulation able to generate completely starless \emph{dark galaxies} with substantial {\HIsc} masses, $M_{\rm {\HI}}>10^{4-5}\,M_{\odot}$, (Fig.~\ref{fig: mhalo_mHI} and Fig.~\ref{fig: Cloud9_mHI_rMW}). Thanks to its physically motivated SF recipe, {\nivarialg} is likely to provide the most reliable predictions for \emph{dark galaxies} with sufficient {\HIsc} to be detectable. Indeed, while all \emph{dark galaxies} remain completely starless in {\nivarialg}, the low SF density threshold adopted in {\hestia} allows some \emph{dark galaxies} to form a small number of stars.

The study of \emph{dark galaxies} is currently a highly active field of research, driven by the increasing availability of both observational data and theoretical predictions. Improvements in observational techniques are expanding the number of \emph{dark galaxies} candidates, while advances in numerical simulations, now able to reproduce many of the key processes governing galaxy formation and evolution, are proving essential for interpreting these systems. Together, these developments are pushing the field forward and bringing us closer to a robust confirmation of genuine \emph{dark galaxies}. Such objects would serve as powerful cosmological probes of the \lcdm\ model, and their detection would provide strong support for it.
We expect upcoming observations with FAST, along with other radio telescopes, to aid in the detection of {\HIsc} sources harboured within a DM halo. Future work should focus on measuring the gas metallicity within \emph{dark galaxies}, to determine whether truly starless systems indeed contain primordial, metal-free gas, and to assess whether any past SF episodes may have occurred.

\begin{acknowledgements}
We thank Julio Navarro, Duncan Forbes, Jonah Gannon and Marco Monaci for useful discussions. We thank the anonymous referee for the helpful comments and suggestions for this work.  ADC, GGB, and  ACS acknowledge financial support from the Spanish Ministry of Science and Innovation (MICINN), Consolidación Investigadora program, CNS2023-144669, project “Tiny” (PI A. Di Cintio), and the 2024 call “Proyectos de Generación de Conocimiento”, grant number PID2024-160009NA-I00, proyecto “INGENIO”. SC acknowledges funding from the State Research Agency (AEI-MICINN) under the grant with reference PID2023-149139NB-I00.  EAG acknowledges support from (\textsc{AEI-MICINN}) and the European Social Fund (\textsc{ESF+}) through a FPI grant PRE2020-096361. SCB acknowledges financial support from the (MICINN) through RYC2022-035838-I,  PID2021-128131NB-I00 (CoBEARD) and CNS2022-135482 projects. AN is supported by the applied research and innovation project (SOL2024-31834, PI Andrea Negri), co-financed by the EU - Ministry of Finance and Public Service - European Funds - Junta de Andalucía – Consejería de Universidad, Investigación e Innovación. AK is supported by project PID2024-156100NB-C21 financed by MICIU /AEI/10.13039/501100011033 / FEDER, UE and further thanks Fontaines DC for hurricane laughter. This research is also co-funded by the European Union (Widening Participation, ExGal-Twin, GA 101158446 “UNDARK”, GA 101159929). The {\nivarialg} simulations have been run using \emph{LaPalma} supercomputer, project \textit{can43} (PI A. Di Cintio). The authors acknowledge the contribution of the IAC High-Performance Computing support team.
\end{acknowledgements}

\bibliographystyle{aa}
\bibliography{biblio} 

\onecolumn
\appendix

\section{Evolution of the density of the environment} \label{appendix_1}

We present in Fig.~\ref{fig: dens_envirn_median} the median densities of the environment around dark and bright galaxies for each simulation individually. In this figure, \emph{dark galaxies} are represented as dotted-dashed lines and bright galaxies as solid lines. Here, the peaks are  owing to pericentric passages of galaxies that become satellites of other galaxies at different times, and therefore the density of their environment rises steeply because of the presence of a larger galaxy. Note that the cyan dotted-dashed line refers to the only \emph{dark galaxy} found in the {\hestia} run 37\_11, which is currently a satellite galaxy (see Table~\ref{tab: sample}).

    \begin{figure*}[h]
    \centering
    \includegraphics[width=9cm]{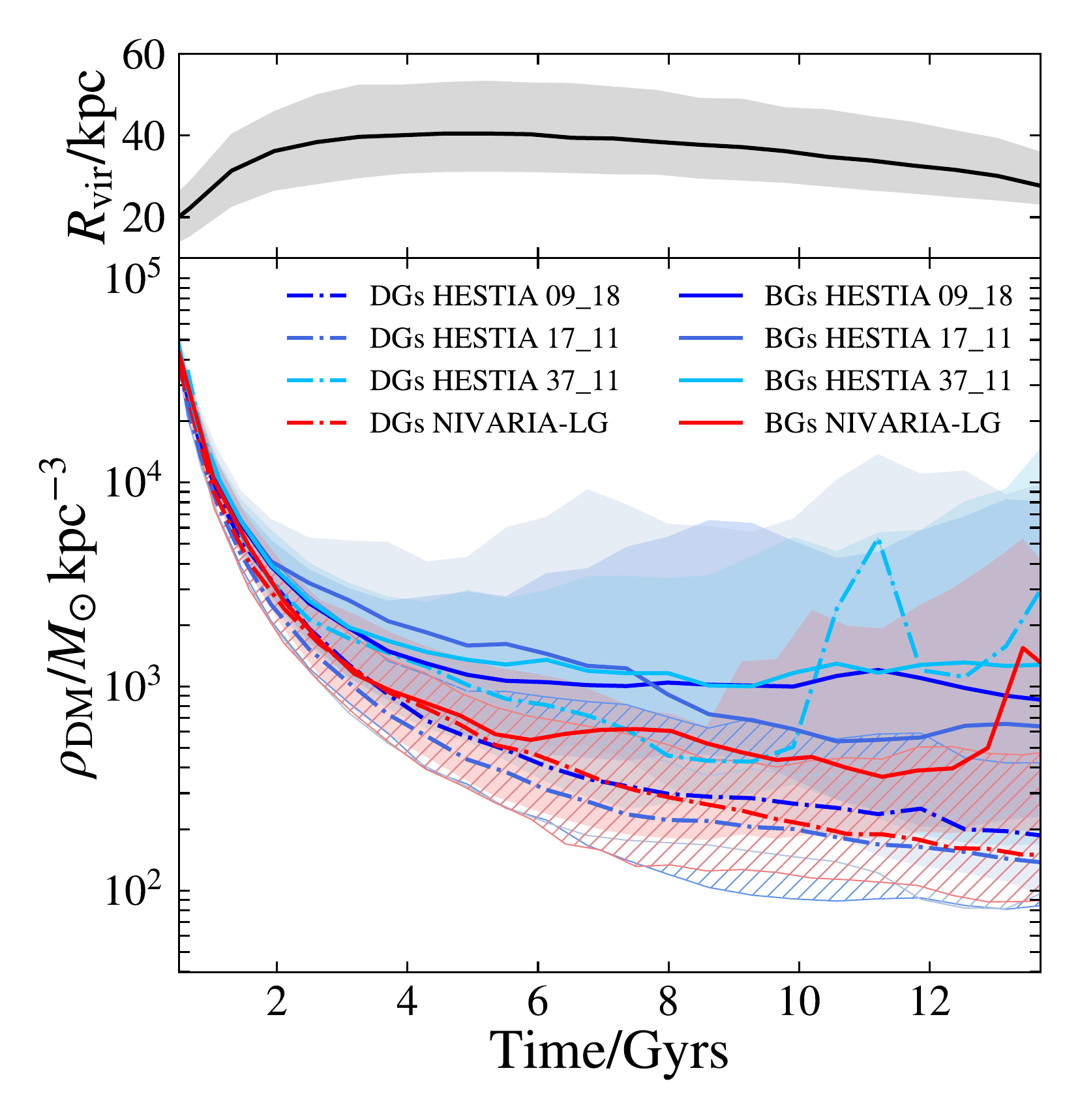}
    \includegraphics[width=9cm]{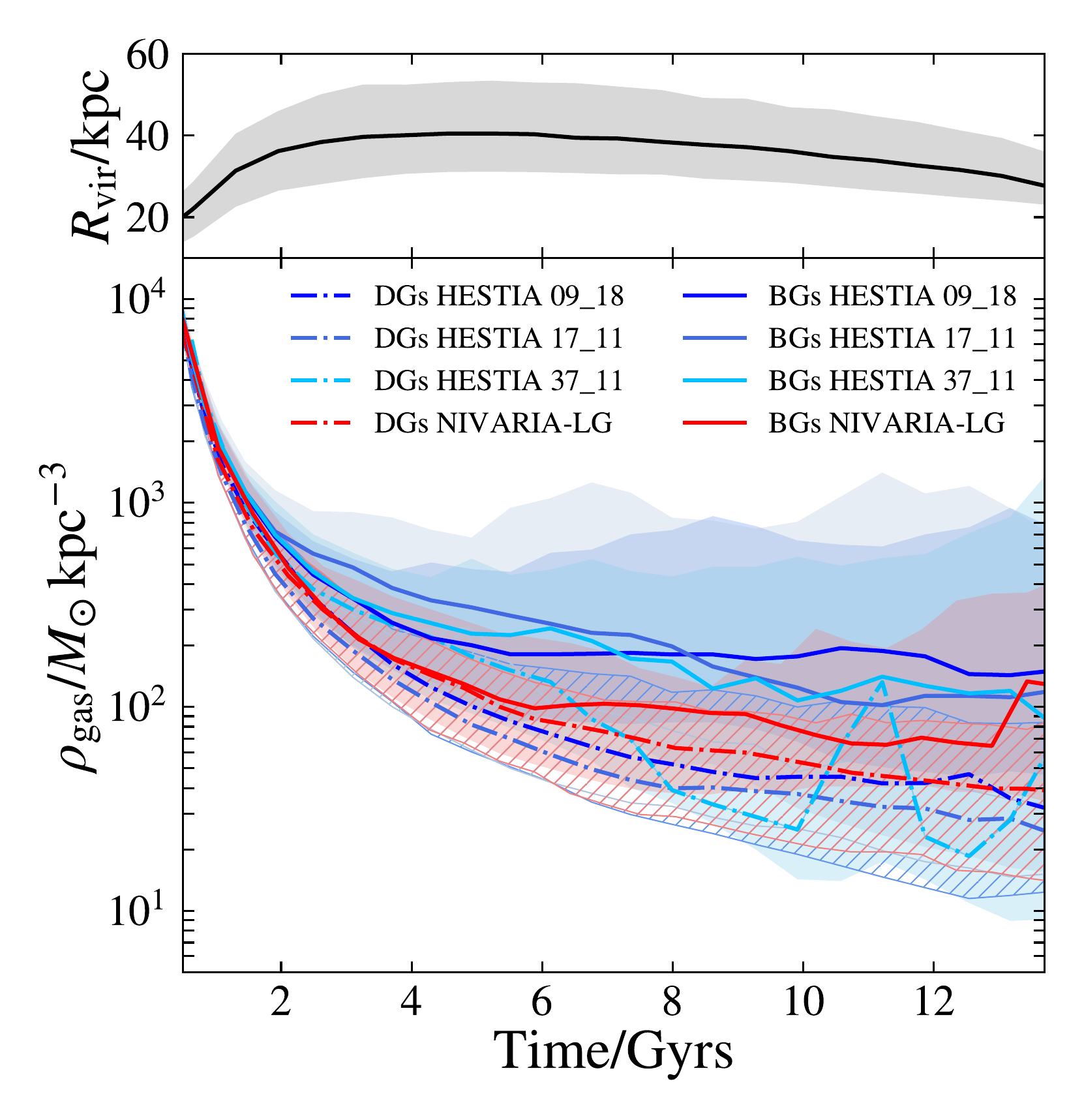}
   \caption{Evolution of median DM (\textit{left bottom panel}) and gas (\textit{right bottom panel}) density of the environment around dark and bright galaxies. \emph{Dark galaxies} are shown as dotted-dashed lines, and bright galaxies as solid lines. Each colour represents a different simulation, blue shades for {\hestia} and red for {\nivarialg}. The shaded and striped backgrounds are the regions limited by the $16^{\rm th}$ and $84^{\rm th}$ percentiles of the distributions. The environment is defined as a spherical shell between 1 $R_{\rm vir}$ and $7\,R_{\rm vir}$ of each halo at each time. The evolution of the median $R_{\rm vir}$ of all dark and bright galaxies is shown in the \textit{top} panels as a black solid line, along with the $16^{\rm th}$ and $84^{\rm th}$ percentiles as a shaded region.}
          \label{fig: dens_envirn_median}%
    \end{figure*}

\twocolumn
\section{FAST detection limits} \label{appendix_2}

We follow the procedure presented in \cite{2018RAA....18....3Li}, done to obtain the integration time and number of detectable galaxies. 

Firstly, we need the thermal noise associated to the instrument (\citealt{2008MNRAS.383..150Duffy}):

\begin{equation}
    \sigma_{\rm{noise}} = \sqrt{2} \dfrac{kT_{\rm{sys}}}{A_{\rm{eff}}} \dfrac{1}{\sqrt{\Delta\nu t}},
    \label{eq: thermal_noise}
\end{equation}

\noindent where $k = 1380\,\rm{Jy \, m^2 \, K^{-1}}$ is the Boltzmann constant, $T_{\rm{sys}}$ is the temperature of the system, $A_{\rm{eff}}$ is the aperture efficiency of the telescope, $\Delta \nu$ is the frequency bandwidth, and $t$ is the integration time.

Once the thermal noise is obtained, we can compute the minimum observed flux limit at which a galaxy can be detected at $z = 0$. This threshold is given by the following expression (\citealt{2008MNRAS.383..150Duffy}):
 
\begin{equation}
    S_{\rm{lim}} = \rm{(S/N)}\, \sigma_{\rm{noise}}, 
    \label{eq: flux_limit}
\end{equation}

\noindent where S/N is the signal-to-noise ratio. 

Now, the minimum {\HIsc} mass that a galaxy needs to have in order to be detectable by FAST at $z = 0$ can be directly derived as follows (e.g. \citealt{1975gaun.book..309Roberts}; \citealt{2017PASA...34...52Meyer}):
\begin{equation}
    \dfrac{M_{\rm{{\HI}}}}{M_{\odot}} = 2.35 \times 10^5 \left(\dfrac{d}{\rm{Mpc}}\right)^2 \left(\dfrac{S_{\rm{lim}}}{\rm{Jy}}\right) \left(\dfrac{\Delta V_{0}}{\rm{km \, s}^{-1}}\right), 
    \label{eq: HImass}
\end{equation}

\noindent where $d$ is the luminosity distance, and $\Delta V_0$ is the velocity linewidth.

We adopt the same parameters as in \cite{2018RAA....18....3Li}. This is an aperture efficiency of $A_{\rm{eff}} = 5 \times 10^4\,\rm{m}^2$ for the spherical main reflector of FAST; a system temperature of $T_{\rm{sys}} = 25\,\rm K$ , correspondent to the 21 cm line at 1.4 GHz within the $L$-band; a frequency bandwidth of $\Delta \nu = 10^{6}$ Hz; a velocity linewidth of $\Delta V_0 = 200\,\rm{km \, s^{^-1}}$; and a signal-to-noise ratio of $\rm S/N= 6$. Lastly, we set $z = 0$, assume 1 hour of observing time, and obtain $M_{\rm{{\HI}}}$ for two different distances, namely $\sim 1\,\rm Mpc$ and $\sim 2.5\,\rm Mpc$. This setup results in minimum masses of $M_{\rm{{\HI}}} \, (1 \, \rm{Mpc}) = 4.6 \times 10^{3}\,\textit{M}_{\odot}$, and $M_{\rm{{\HI}}} \, (2.5 \, \rm{Mpc}) = 2.9\times 10^{4}\,\textit{M}_{\odot}$ for galaxies within $1\,\rm Mpc$ and $2.5\,\rm Mpc$\footnote{Note that $d$ in Eq.~\ref{eq: HImass} refers to luminosity distance from Earth, whilst our distances are measured with respect to the {\ahf} centre of the MW analogues.}, respectively.

\end{document}